\documentclass[onecolumn]{aastex61}
\usepackage{amsmath,booktabs,graphicx,subfigure,amssymb,latexsym,graphics,epsfig,exscale,url,booktabs}
\usepackage[flushleft]{threeparttable}
\usepackage{epstopdf}
\usepackage{gensymb}
\usepackage[section]{placeins}
\usepackage[colorinlistoftodos]{todonotes}
\usepackage{float}
\usepackage{chngcntr}
\counterwithin*{footnote}{section}
\usepackage[english]{babel}
\usepackage[utf8x]{inputenc}
\begin{document}

\title{Protoplanetary Disks in $\rho$ Ophiuchus as Seen From ALMA}
\correspondingauthor{Erin G. Cox}
\email{egcox2@illinois.edu}

\author{Erin G. Cox}
\affil{Department of Astronomy, University of Illinois at Urbana-Champaign,
1002 W. Green St.,
Urbana, IL 61801, USA}

\author{Robert J. Harris}
\affiliation{National Center for Supercomputing Applications, Urbana, IL 61801} 

\affil{Department of Astronomy, University of Illinois at Urbana-Champaign,
1002 W. Green St.,
Urbana, IL 61801, USA}

\author{Leslie W. Looney}
\affil{Department of Astronomy, University of Illinois at Urbana-Champaign,
1002 W. Green St.,
Urbana, IL 61801, USA}

\author{Hsin-Fang Chiang}
\affiliation{National Center for Supercomputing Applications, Urbana, IL 61801}

\author{Claire Chandler}
\affiliation{National Radio Astronomy Observatory, P.O. Box O, Socorro, NM 87801, USA}

\author{Kaitlin Kratter}
\affiliation{Department of Astronomy and Steward Observatory, University of Arizona, Tucson, AZ 85721 USA}

\author{Zhi-Yun Li}
\affiliation{Department of Astronomy, University of Virginia, Charlottesville, VA 22903, USA}

\author{Laura Perez}
\affiliation{Max Planck Institute for Radio Astronomy, Bonn, Germany 53121}

\author{John Tobin}
\affiliation{Homer L. Dodge Department of Physics and Astronomy, Norman, OK 73019, USA}
\affiliation{Leiden Observatory, Leiden University, P.O. Box 9513, 2000-RA Leiden, The Netherlands}

\begin{abstract}
We present a high angular resolution ($\sim 0.2\arcsec$), high sensitivity ($\sigma \sim 0.2$ mJy) 
survey of the 870 $\mu$m continuum emission from the circumstellar material around 49 pre-main sequence stars in 
the $\rho$ Ophiuchus molecular cloud. Because most millimeter instruments have resided in the northern hemisphere,
this represents the largest high-resolution, millimeter-wave survey of the circumstellar disk content of this cloud.  
Our survey of 49 systems comprises 63 stars; we detect disks associated with 29 single sources, 11 binaries, 3 triple systems and 4 transition disks. We present flux and radius distributions for these systems; in particular, this is the first presentation of a reasonably complete probability distribution of disk radii at millimeter-wavelengths. We 
also compare the flux distribution of these protoplanetary
disks with that of the disk population of the Taurus-Auriga molecular cloud. We find that disks in binaries are both 
significantly smaller and have much less flux than their counterparts around isolated stars.
 We compute truncation calculations on our binary sources and find that 
these disks are too small to have been affected by tidal truncation and posit some explanations for this. 
Lastly, our survey found 3 candidate gapped disks, one of which is a newly identified transition disk with no signature of a dip in infrared excess 
in extant observations.
\end{abstract}

\section{Introduction}  \label{sec:intro} 
In recent years, there has been an explosion in the detections of extra-solar planets.  As of early 2017, there are nearly 3000 exoplanets confirmed 
and another 2500 candidate exoplanets (e.g., exoplanets.org).  These planets show a great diversity of properties including masses, sizes, and 
architectures.  In fact, many of these systems have planets that are unlike our solar system, such as super Earths, hot Jupiters, or hot Neptunes 
\citep{ch13}. The diversity of the planet population is likely some combination of differences in the 
initial conditions during the evolution of the circumstellar disk in which the planetary system forms and the necessary random
interactions or scattering events during the planetary growth process \citep{bi11,bi15}.  To better understand the origin of the planet diversity, we therefore 
need to explore the inherent diversity in the circumstellar disks around young stellar objects
(hereafter, YSOs). By directly observing the environments in which young planetesimals are expected to 
form, we can characterize the initial conditions of these other worlds. 

To explore these early conditions, we must observe the protostar at the 
evolutionary phases that likely have the largest
impact on planet evolution.   While the exact phase is still unknown, the protostar must have evolved
to the point where a large mass reservoir, i.e., a protoplanetary disk, surrounds the star. A protostar's evolutionary path can be
divided into 4 parts- Class 0 -- III \citep[e.g.,][]{la87,an93,du14}. During the initial collapse, i.e. the Class 0 phase,
the protostar is engulfed in a large envelope full of nascent dust and gas. 
By the Class I phase, most of the envelope material has been funneled onto
the central protostar through a circumstellar disk. During the Class II phase, the protostar no longer has its nascent envelope
surrounding it, and the majority of the circumstellar material is in a large disk. Lastly, during the final phase of 
the protostar, Class III, the 
protostar has essentially accreted all of its final mass,
leaving a very tenuous (if any) circumstellar disk left \citep[e.g.]{an05,an07}.

It is well known that planets form in the disk surrounding forming protostars, and it is commonly
thought that most of planet formation happens during the Class II phase of evolution. This is
due to the fact that, by this time, the majority of the remaining gas and dust are surrounding the central protostar
in a disk, allowing a large reservoir for planetesimals to form and evolve. While there is overwhelming indirect evidence for
planet formation in disks, direct imaging of forming protoplanets has been scarce, with few examples in the literature 
(LkCa 15, \citealt{ki12}; FW Tau; ROXs 12; ROXs 42B, \citealt{kr14}).
However, recent Atacama Large Millimeter/submillimeter 
Array (ALMA) observations of protoplanetary disks are beginning to reveal 
likely indicators of ongoing planet formation, such as the gaps in the millimeter disks of HL Tau \citep{al15}, a Class I/II protostar,  and of TW Hydra \citep{an16}, a Class II protostar. 

$\rho$ Ophiuchus is an ideal laboratory for studying star and planet formation for several reasons.
First, it is relatively close ($d \sim 137$ pc; \citealt{or17}); second, it is relatively young (between 0.5 - 2 Myr; \citealt{wi08});
finally, it has a large number of confirmed/candidate members ($\gtrsim 300$; \citealt{wi08}).
Despite these advantages, there are few millimeter-wave studies of its disk population that are representative of the disk content of Oph. One reason for this is that the stellar population is not well-characterized or studied: 
with $A_v$ ranging from 1 to 100 across the cloud, an accurate/representative stellar census has not been possible to date, despite many optical/IR surveys 
of different parts of the cloud (see, e.g, \citealt{ba03} and references therein), making connection to host star properties difficult. 
Another reason is that Oph lies far in the southern hemisphere, making it somewhat challenging to observe with northern instruments.

Of the few large-scale surveys toward Oph, most have been done with 
single-dish telescopes, and thus are potentially confused by cloud contamination, companion stars, etc. 
The first studies of the Oph cloud core \citep{an90,le91} 
showed an abundance of millimeter/centimeter-bright, deeply embedded objects residing in the dense core. 
Subsequent systematic studies of both the cloud core 
and surrounding regions \citep{an94,an07} demonstrated that millimeter flux tends to decline with class, signifying circumstellar 
mass depletion during evolution (either through accretion 
or outflow or dispersion, by, e.g., photo-evaporation), and also that the millimeter spectral index tends to decline as well, most likely 
indicating grain growth in the circumstellar dust \citep[e.g.,][]{ri10}.

Subsequent work at high resolution with both the Submillimeter Array (SMA) and ALMA have yielded more details by probing the disk 
structures in Oph at sub-arcsecond resolution. These studies have, however, focused principally on
either the detailed structure of the transitional disk population of Oph 
\citep{an09,an10,pe14} or other special (i.e., bright) objects \citep{pe12,sa14}. 
Despite these studies of special sub-populations of Oph disks, there has, to 
date, been no systematic study at high-resolution ($<$ 0.2\arcsec) of the disks of the $\rho$ Ophiuchi cloud complex.
In this article, we present the results from our ALMA 870 $\mu$m survey of $\sim$50 evolved
disks in $\rho$ Ophiuchus. 
\section{Sample Selection} \label{sec:sam}
One of the main goals of this program is to observe the compact disk dust emission toward a large sample of 
sources that does not have the inherent biases of previously known millimeter flux detections. To achieve such a large 
 and representative sample of sources, we used the Spitzer c2d catalog of YSO candidate sources in $\rho$ 
Ophiuchus \citep{ev03}, which 
requires S/N $\geq$ 3 in all the 4 IRAC bands and the 24 $\mu$m MIPS band. This criteria yields 297 protostellar sources.
To increase the likelihood of detectable circumstellar mass (i.e. long wavelength dust emission), 
we narrowed the sample to sources with 
70 $\mu$m MIPS band detection S/N $>$ 2.  This requirement removed 
mostly the older source population (e.g. Class III objects
based on SED fitting between 2-24 $\mu$m) and other sources that
have low-mass disks due to other factors (i.e. environment, system mass, etc.), including 18 Flat and 10 Class I sources, 
which left 64 sources.

Finally, as this project is focusing
on the more evolved sources without significant envelope emission, we also 
removed the sources that were known embedded sources from Young et al. (2006). This resulted in a total of 50 sources in our sample.  
These sources were then compared to Herschel PAC continuum maps at 70 and 100 $\mu$m to verify that the sources 
all had far-infrared emission. While doing this, it was realized that one of the sources was a clear galaxy (J163524.3-243359) 
and another was offset by exactly 1 arcminute (J162646.4-241160), which was likely a typo in the c2d catalog and is now corrected.  
The final source list of 49 sources with their YSO class from the c2d catalog are given in Table 1. 

\begin{table}
\begin{center}
\begin{threeparttable}
\setlength\tabcolsep{1.0pt}
\caption{Multiplicity of target sources}
\begin{tabular}{llcc|llccc}
\toprule
\multicolumn{4}{c}{Known singles} & \multicolumn{5}{c}{Known binaries} \\
\cmidrule(r){1-4}  \cmidrule(r){5-9} 
Field name   & Alt name  & & Ref &  Field name & Alt name & Separation (arcsec) & PA ($^\circ$)  & Ref  \\
\hline
ROph3  & 	IRAS 16201-2410          & & e & ROph2  &  V 935 Sco& 0.02  & \ldots& d  \\
ROph4  &           & & e & ROph5       &  WSB 19   & 1.49  & 262.9 & a \\  
ROph8 		 & DoAr 25   & & a,c& ROph6 		 & DoAr 21   & $\sim 0.005$  & \ldots & g\\
ROph9 		 & El 24     & & a,b,c  &  ROph7       &  DoAr24 E& 2.03       & 150   & a \\
ROph10 		 & GY 33     & & a  &  ROph12 & WSB 40   & 0.017      &  \ldots &  d  \\
ROph14 		 & GY 211 	 & &  c  &  ROph21      & SR 9     &  0.638     & 353.3 & a \\
ROph15 		 & GY 224 	 & &  a,b  &  ROph26      & ROXs 42C &  0.277    &  151  & a \\
ROph16 		 & GY 235	 & &  a  & ROph27      &  WSB 71  &  3.56      &  35.0 & a\\ 
ROph17 		 & GY 284 	 & &  a  & ROph32	  & WSB 74 &  $\lesssim 0.043$ & \ldots & e  \\ 
ROph18 		 & YLW 47	 & & a,c & ROph33     &  DoAr 51 &  0.784    &  79.3 & a \\
ROph19 		 &DoAr 33  & &a,c  & ROph34      &  L1689-IRS7 & 7.56 & 334.9 & h\\ 
ROph20 		 & GY 314 	 & &a,c   & ROph36       &           &   0.025  & \ldots      & d \\
ROph22 		 & SR 20 W 	 & &a,c  &    ROph45     &  IRS 54   & 7.17  &      323.1 & b \\
ROph24 		 & WSB 63 	 & &a,c  &\multicolumn{5}{c}{Known triples}  \\ \cmidrule(r){5-9}   
ROph25 		 & WSB 67 	 & &a   &   Field name & Alt name & Separation  (arcsec) & PA($^\circ$) & Ref  \\ \cmidrule(r){5-9}
ROph29 		 & DoAr 44   & &a,c,f   &   ROph11     &  WSB 38 Aa-Ab   &   0.098 &  24.2 & a \\ 
ROph35 		 & 	Haro 1-17& &a  &          & WSB 38 Aab-B &  0.577  &  105.4 & a \\
ROph40  &           & & a &  ROph13      &  SR 24 Aab   & 0.197   &  84   & a \\      
ROph41 		 & WL6 		 & &a,b  &  & SR 24 Aab-B  & 5.065   & 349  & a \\
ROph42       & GY 312    & &b   &   ROph23      & SR 13 Aa-Ab    & 0.013     & ...$^{\dagger}$ & a \\
ROph43       &           & &b    &             & SR 13 Aab-B  & 0.399   &  96   & a \\
ROph44        &  GY 344        & &b &             ROph31      &  L1689-IRS5 A-Bab & 3.0  &  241 & a\\  
ROph46 		 & WSB 60 	 & &a  & &  L1689-IRS5 Ba-Bb &  0.14  & 84.4 & a \\
ROph48 		 & IRS 63 	 & &a    &  \\
ROph50 		 & Haro 1-11  & &a,c    &  \\
   \multicolumn{4}{c}{No data on companion objects } \\ 
   \cmidrule(r){1-4}   
  Field name & Alt name  & Field Name  & Alt Name  \\ 
 \cmidrule(r){1-4}  
 ROph1 &    &   ROph39  &  \\
  ROph28&    &   ROph47  &  \\
ROph30&    &   ROph49  & \\
   ROph38& WSB 82  &           & \\

\bottomrule
\end{tabular}
\end{threeparttable}
    \begin{tablenotes}
      \small
      \item $^{\dagger}$  binary orbits with a period of $\sim$ years, so the position angle depends sensitively on observation epoch.
      \item Reference key: [a] \citealt{ra05}, [b] \citealt{du04} , [c] \citealt{ch15}, [d] \citealt{rr16}, [e] \citealt{ko16}, [f] \citealt{wi16}, [g] \citealt{lo08}, [h] This work
    \end{tablenotes}
\end{center}
\label{tab:multtab}
\end{table}

Because we select for sources that have infrared excesses in each of the IRAC and MIPs bands, we preferentially observe sources with a substantial disk reservoir. Since mass estimates at longer wavelengths are less affected by optical depth than those at shorter wavelengths, we 
attempt to quantify this bias by computing model disk fluxes at 70 $\mu$m and comparing them 
to the observed MIPS 70 $\mu$m fluxes in our sample. To do this, we assume the standard analytic prescription for a viscously-evolving, geometrically-thin disk \citep{ly74,ha98}, a radial power-law in temperature, and a power-law in frequency for the total (i.e., gas + dust) opacity \citep{hi83}, i.e.

\begin{eqnarray*}
\Sigma(r) &\propto& \left(\frac{r}{r_c} \right)^{-\gamma} \exp{-(\frac{r}{r_c})^{2-\gamma}}\\
T(r) &=& T_{\mathrm{1 au}} \left(\frac{r}{\mathrm{1 AU}}\right)^{-q} \\
\kappa_{\nu} &=& 0.03 \left( \frac{\lambda}{870 \mu\mathrm{m}}\right)^{-\beta}\mathrm{cm}^2\mathrm{g}^{-1}\\
\end{eqnarray*}
with $T_{\mathrm{1 AU}} = 280$ K, $q = 0.5$, $\beta = 1$, $r_c = 100$ au, and $\gamma = 1$. These values and expressions are roughly appropriate for these disks as observed in the (sub)-millimeter (e.g., \citealt{hu08,an09,an10}), although their applicability to the mid/far-infrared is uncertain.   
The median uncertainty for the c2d survey of Ophiuchus at 70 $\mu$m is approximately 25 mJy, so our 70 $\mu$m selection criteria selects sources with fluxes in excess of $\sim 50$ mJy at 
70 $\mu$m. Using these relations, we estimate that our sources all have $\gtrsim 0.2 - 1$ 
Jupiter mass worth of circumstellar material (gas + dust), depending on the exact values for the quoted values above, as well as the relatively uncertain gas-to-dust ratio used for the computation
of the opacity.

\begin{table}
\begin{center}
\begin{threeparttable}
\setlength\tabcolsep{2.0pt}
\caption{Transition vs. non-transition disks}
\begin{tabular}{llr|llcc}
\toprule
\multicolumn{3}{c}{Not transition disks based on N/FIR colors} & \multicolumn{4}{c}{Objects with transition disk colors or millimeter cavities} \\
\cmidrule(r){1-3}  \cmidrule(r){4-7} 
Field name   & Alt name  &IR band/ref &  Field name & Alt name & band/ref & True disk or binary/ ref  \\
\hline

ROph1 &        &   N [a]  & ROph2 & V 935 Sco & N [a] & CB [2]\\
ROph5 & WSB 19 &  N [a]   &  ROph 3$^{\dagger\dagger}$  &	IRAS 16201-2410   &  N,S [c,f]    &  T [3] \\
ROph7 & DoAr 24 E& N [a]  & ROph4 &           & N [a] & T [3] \\
ROph9 & El 24 &   N,M [a,b]  & ROph6 & DoAr 21   & N,M [a,b] & CB [4]\\
ROph10& GY 33 &   N [a] & ROph8 & DoAr 25   & N,M [a,b] & T [2,3]\\
ROph14& GY 211&   N [a]  & ROph11& WSB 38  & N [a] & T [3] \\
ROph15& GY 224&   N [a]   &  ROph12& WSB 40    & M [b]  & T [2] \\
ROph16& GY 235&   N [a]  & ROph13& SR 24     & M,S [b,d]  & T [2]\\
ROph25& WSB 67&  N [a]  & ROph17& GY 284    & M [b]  & T [7]   \\
ROph27& WSB 71&  N [a]    & ROph18& YLW 47    & M [b]  & T [5] \\
ROph28&       &   N [a]  &  ROph19& DoAr 33   & N [a]  & T [2,3,5]\\
ROph30&        & N [a]  & ROph20& GY 314    & M [b]  & T [5] \\
ROph31& L1689-IRS5& N [a] &  ROph21& SR 9      & N,M [a,b]  & T [3,5]\\
ROph33& DoAr 51&   N [a] & ROph22& SR 20 W   & M [b]  & T [5] \\
ROph34& L1689-IRS7&N [a]& ROph23& SR 13     & M [b]  &  CB [1] \\
ROph35& Haro 1-17& N [a]  &  ROph24& WSB 63    & N [a]  & T [2,3,5]\\
ROph39&        &  N [a] & ROph26& ROXs 42C  & N [a]  & T [5]  \\
ROph40&   ISO-Oph 51     &  N,M [a,b]  &ROph29$^\dagger$&  DoAr 44 & S [e] & T [2,5,6]\\ 
ROph41&  WL 6  &  N [a]  & ROph32&  WSB 74   & N [a]  & CB [3] \\
ROph42& GY 312 &   N [a]  & ROph36&           & N [a]  & CB [2]\\
ROph43&        &  N [a]   &ROph38& WSB 82 &  S [f]  & T [7]  \\
ROph44&  GY 344 & N [a]   &ROph46& WSB 60    & M [b]  & T [7] \\
ROph45& IRS 54 &   N [a]   &ROph50& Haro 1-17& N [a]   & T [7]\\
ROph47&        &   N [a]    &\\
ROph48& IRS 63 &   N [a]  &\\
ROph49&        &   N [a]  &\\
\bottomrule
\end{tabular}
\end{threeparttable}
    \begin{tablenotes}
      \small
      \item Abbreviation key: N - near/mid-infrared colors; M - mid/far-infrared colors; S - (sub)mm-wave imaging of cavities; T - no indication of interloping circumbinary disk, CB - indication that disk is circumbinary, not transitional.
  
  \item $\dagger$: this source was classified as a pre-transitional
  disk by \citealt{es10} but did not meet the color criterion to be a transitional disk according to \citealt{ci10}. We treat it as a non-transition disk here, for consistency.
  \item $\dagger\dagger$: this source was classified as a transitional disk on the basis of \textit{Spitzer} IRS spectra by \cite{fu09} but, as with ROph 29, the colors did not meet the criteria of \citealt{ci10}. 
 \item Reference key: [a] \citealt{ci10}, [b] \citealt{re15}, [c] \citealt{fu09}, [d] \citealt{an05b}, [e] \citealt{an09}, [f] this work, [1] \citealt{ra05} [2] \citealt{rr16} , [3]  \citealt{ko16}, [4] \citealt{lo08}, [5] \citealt{ch15}, [6] \citealt{wi16},  [7] assumed transition based on lack of data.
    \end{tablenotes}
\end{center}
\label{tab:multtab2}
\end{table}

Out of the 49 targets selected, 12 -- ROph 2, 4, 6, 8, 11, 19, 21, 24,  26, 32, 36, and 50 -- were identified by \cite{ci10} 
to be candidate transitional disks on the basis of
\textit{Spitzer} near-/mid-infrared colors, eleven sources -- ROph 6, 8, 12, 13, 17, 18, 20, 21, 22, 23, and 46 -- 
were identified by \cite{re15} on the basis of \textit{Spitzer}/\textit{Herschel} mid-/far-infrared colors. 
Of these nineteen total transitional disk	candidates, three -- ROph 2, 12, and 36 -- were	discovered by \cite{rr16} to harbor tight stellar binaries ($a \lesssim 3$ au)
whose infrared signature mimicked that of transitional disks, and one -- ROph 32 -- was discovered by \cite{ko16} to be a spectroscopic binary with $a 
\sim 0.6$ au. ROph 6 was found to be a very tight ($\sim 5$ milliarcsecond) binary by \citealt{lo08}; its mid-infrared color is due to the presence of a 
hot ring of material at a large distance from the star, and is most likely not indicative a true transitional disk. This leaves fourteen candidates that are 
`bona fide' transition disks with no evidence of being binary interlopers, see Table \ref{tab:multtab2}.

One caveat to keep in mind for this survey is the impact of unresolved (or unknown) multiplicity in the targets. $\rho$ 
Ophiuchus has been the target of several optical and infrared surveys of varying completeness in the past three decades, both targeting Class I/Flat
\citep[]{ba04,ha04,du04,ha06,du07}
and Class II \citep[]{gh93,ra05} sources. It is known that stellar companions can have a dramatic effect on 
circumstellar material via tidal interactions that preferentially strip
away outer disk material in circumstellar disks and inner disk material in circumbinary disks
\citep[]{ar94,pi05,pi08}. Observationally, truncation manifests as a decreased likelihood of an infrared 
excess in multiple systems as opposed to isolated stars \citep[reflecting absence of an inner disk,][]{ci09,kr11} or decreased millimeter-wave 
continuum emission \citep[reflecting loss of material in either the inner or outer disk,][]
{je94,ha12}. Such signatures, if unrecognized, can bias the results of infrared and millimeter surveys of protoplanetary disks.

To mitigate the effect of this in our sample, we have surveyed the available literature on multiplicity in Ophiuchus to identify which of our targets are multiple systems.  
Unfortunately, the principal surveys we used provide different sensitivities to various separations on the sky and give fairly heterogeneous coverage. 
\citet{ra05} conducted a magnitude limited (K $\leq$ 10.5) speckle imaging survey of 158 principally Class II objects and is sensitive to companions with 
separations between roughly 0.1$\arcsec$ and 6.4\arcsec, down to a contrast ratio of 0.1. \citet{du04,du07} conducted a direct imaging survey of principally class 
I/Flat objects in the mid-infrared with coverage ranging from 0.8$\arcsec$ to 10.0\arcsec. Because most surveys are flux-limited, several of the lower-luminosity sources 
in our sample have not been observed in these surveys. Of the 49 sources in our sample, 10 have not been observed in any available survey (ROph1, 3, 4, 28, 30, 38, 39, 42, 47, 49). Table 1 summaries the multiplicity status of each system in our sample. 

\section{Observations and Data Reduction}  \label{sec:obs}
The ALMA Band 7 observations were taken under proposal 2013.1.00157.S using a continuum only setup to maximize the 
dust continuum sensitivity in two configurations for the snapshot survey.
The lower resolution observations were obtained on 2015 April, 4 in
ALMA configuration C34-1/(2) for $\sim$30 minutes of total time, which
was about 12 seconds of integration time on each source.  The C34-1/(2) configuration baselines ranged from 14 to 356.3 meters with typical
recoverable scale of 8.4\arcsec.
The higher resolution observations were obtained on 2015 July, 24
in ALMA configuration C34-7/(6) for $\sim$47 minutes of total time, which
was about 24 seconds of integration time on each source. 
The  C34-7/(6) configuration baselines ranges from 42 to 1574 meters with 
typical recoverable scale of 2.6\arcsec. In both observations, the 4 continuum bands were centered at 336.5, 338.4, 348.5, and 350.5 GHz.
The quasars J1517-2422 and J1625-2527 were used for bandpass and phase calibration, and Titan
was used for flux calibration.   In this paper, we assume an 
absolute flux calibration uncertainty of $\sim$10\%, but only statistical uncertainties are considered.

The observations were reduced using the Common Astronomy Software Applications (CASA) package \citep{mc07} using the 4.7.0 CASA and ALMA pipeline package.  
Briefly, the pipeline first applies a priori calibrations, such as baseline corrections and phase corrections from water vapor radiometer measurements. 
Then, it conducts a standard interferometric reduction: bandpass calibration, flux calibration, and antenna gain calibrations. 
These calibrations are performed separately for each of the two observations and each science target field was subsequently split out of its parent dataset. 
For the C34-1/(2) configuration observations, we performed a phase only selfcal over the integration time for the sources with $\geq$10$\sigma$ detections to 
improve the S/N in the maps. The C34-7/(6) configuration map S/N was not improved from selfcal, so the selfcal gains were not used.
After the final images were made for each configuration, we checked to ensure that the fluxes measured for each observation were 
consistent; finding that they were, we combined the two datasets and used the combined datasets to produce the images analyzed in this work. 

To produce the final images, we imaged the data using the \texttt{CLEAN} task in CASA. The data were imaged with natural weighting to 
produce a typical resolution of 0.21$\arcsec$ by 0.18\arcsec. Many of our sources are relatively compact (see Fig. 1 and 2) and standard \texttt{CLEAN} was sufficient to 
deconvolve the sources successfully. However, for many of the more extended disks, standard \texttt{CLEAN} left substantial deconvolution 
errors in the residual maps, so we used the multi-scale version of the algorithm to produce images of some of the disks in Figures \ref{fig:singles} and \ref{fig:triples}  as 
well as all of the transition disks in our sample (Figure \ref{fig:trans}). The use of the multi-scale \texttt{CLEAN} algorithm yielded residual maps that were dominated by Gaussian noise.

\section{Results}  \label{sec:res}
\begin{figure*}
\begin{center}
\includegraphics[scale=0.9]{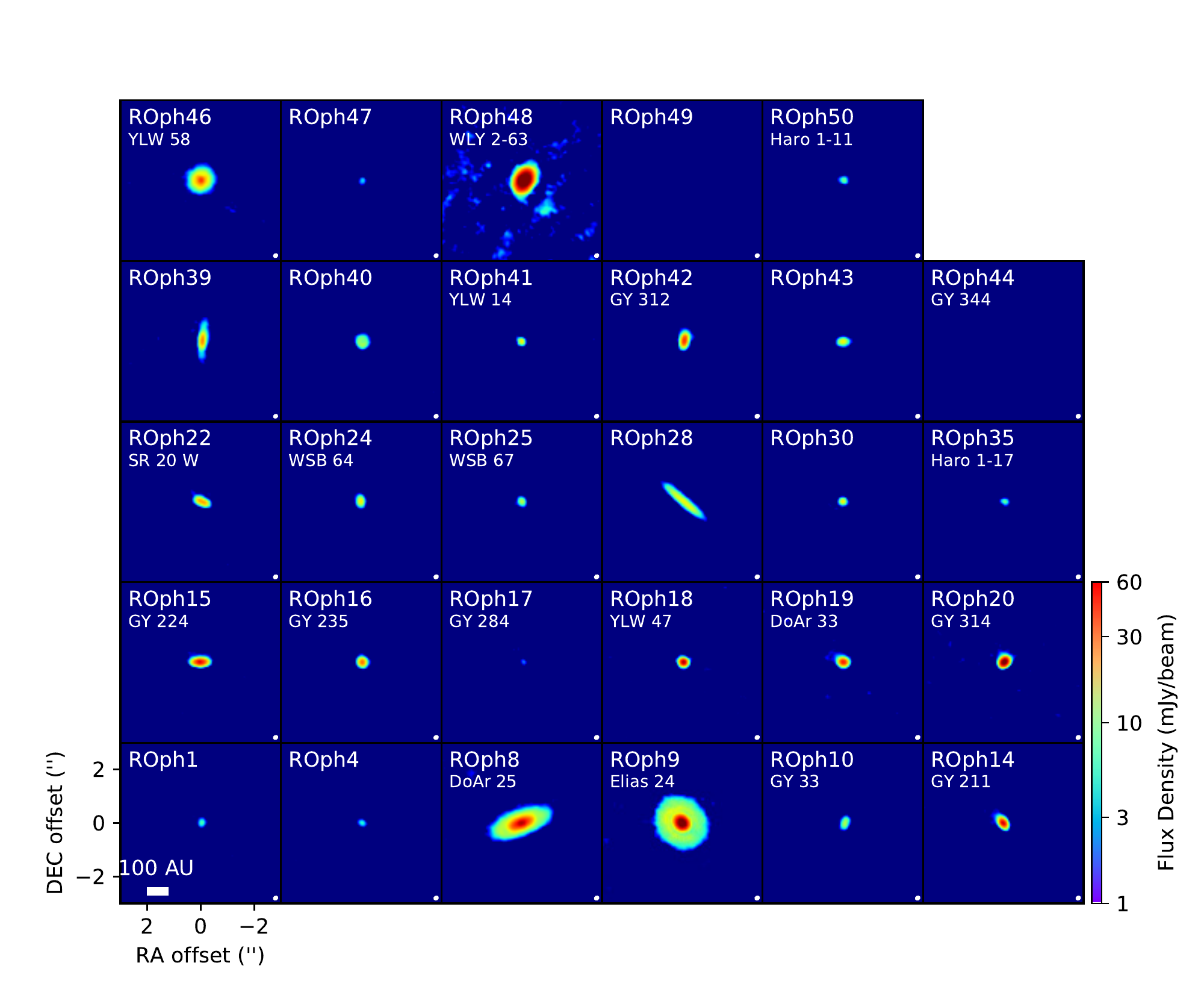}
\end{center}
\caption{Images of the single sources in our sample. The synthesized beam is shown in bottom right corner. ROph1, 3, 4, 28, 30, 39, 42 and 47 have not previously been observed in any survey. Note that the flux scale on the right is a constant scaling for these images.}
\label{fig:singles}
\end{figure*}

\begin{figure*}[htp]
\begin{center}
\includegraphics[scale=0.8]{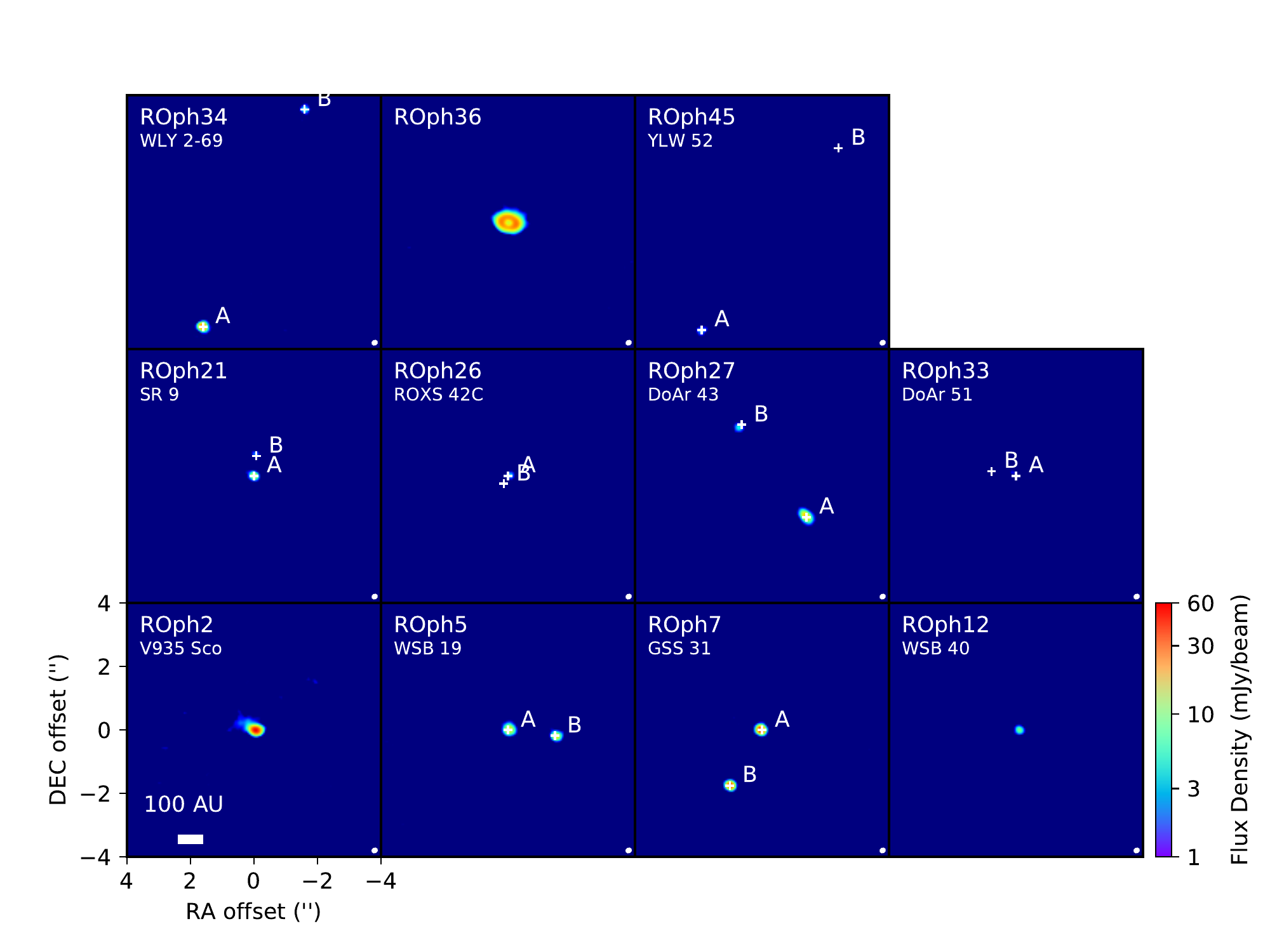}
\end{center}
\caption{Images of the binary sources in our sample. The synthesized beam is shown in bottom right corner. Stellar positions are indicated by white crosses.}
\label{fig:binaries}
\end{figure*}

\begin{figure*}[htp]
\begin{center}
\includegraphics[scale=0.8]{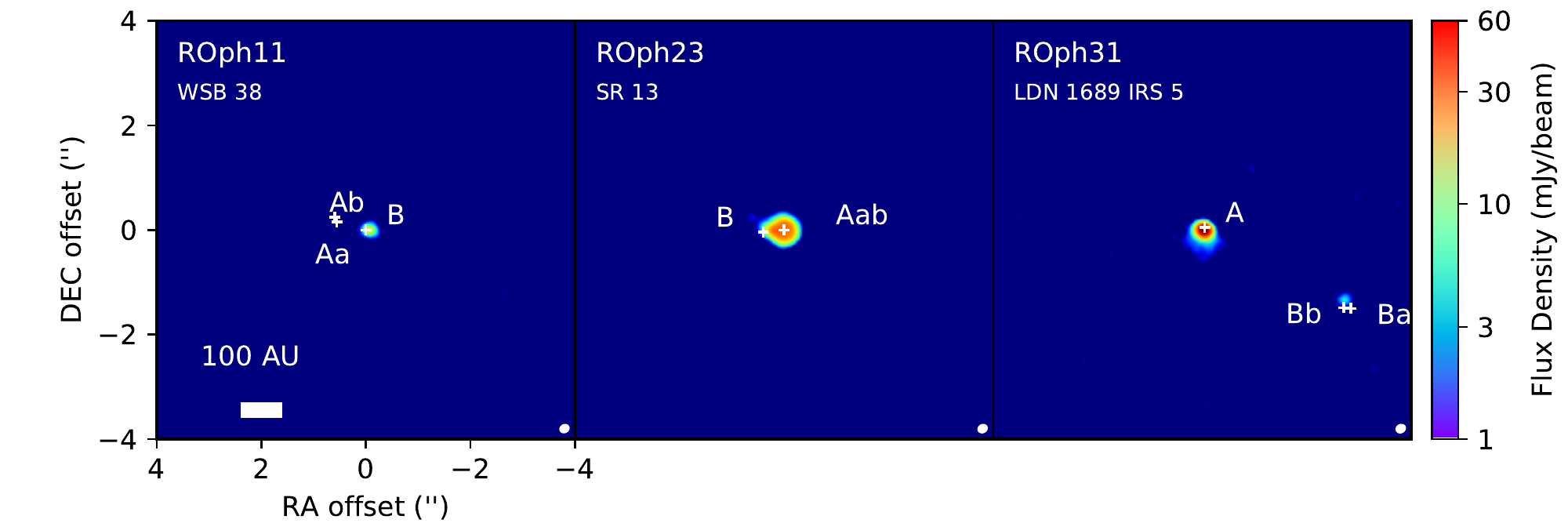}
\end{center}
\caption{Image of the triple systems in our sample. The center panel shows ROph23 with an asymmetrical, circumbinary disk surrounding Aab.
The synthesized beam is shown in the bottom right corner. Stellar positions are indicated by white crosses.}
\label{fig:triples}
\end{figure*}

\begin{figure*}[htp]
\begin{center}
\includegraphics[scale=0.9]{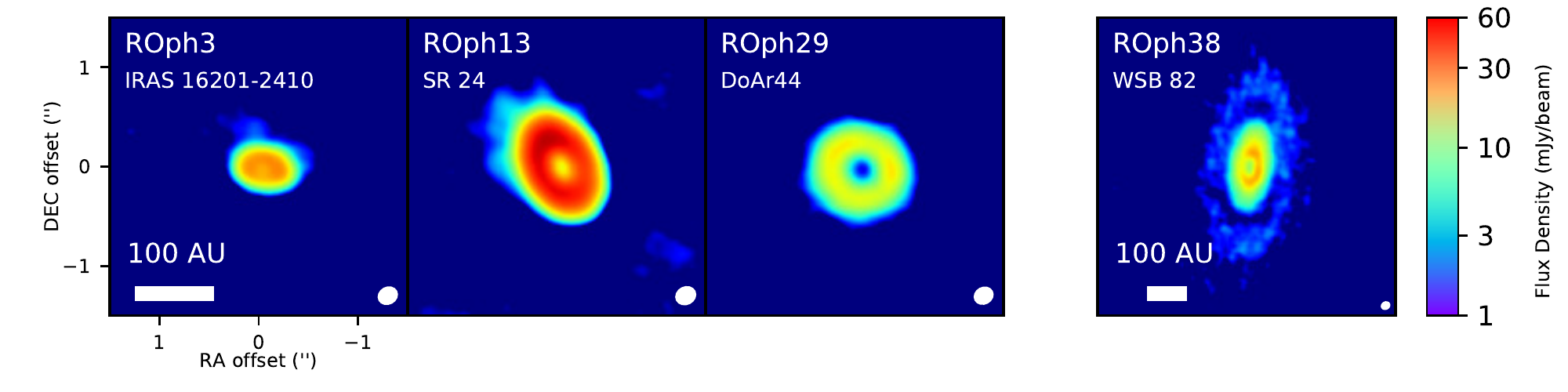}
\end{center}
\caption{Images of 4 transition disks with substantial millimeter cavities in our sample. The source on the right (ROph 38) is a particularly 
large transition disk, with low-level emission gap. 
This source has also not previously been observed in any sub/mm-wave survey. The synthesized beam is shown in bottom right corner.}
\label{fig:trans}
\end{figure*}

This survey provided very well-resolved images of the diverse population of
protoplanetary disks in $\rho$ Oph YSOs. Figures \ref{fig:singles}-\ref{fig:trans} show the different YSOs divided into single sources,
binaries, triple systems and transition disks. The sources that do not have multiplicity information are considered single, 
unless they show evidence of being a transition disk. Since transition disks are separated into
their category, we do not include them in any other categories (i.e., singles or triples).
Each figure uses the same stretch
for flux values, such that the brightest sources show the deepest red color. At a glance,
it is obvious that our sample is not only composed of different types of systems, but
also each type shows great diversity in size, brightness, and flux distribution. 
In the 49 stellar systems that we targeted, there were 63 stars, and disks associated with 13 stars were not detected: 4 around single stars, 4 around components of binaries, and 5 around components of triple systems.
Table \ref{tab:source} summarizes
this information for all sources, including classifications of the YSOs from c2d. The disk sizes and position angles, as well as the peak and integrated fluxes, were estimated by fitting a gaussian in the image plane 
using the CASA task \texttt{imfit}. Disk masses were estimated from the integrated fluxes by assuming optically thin emission and an isothermal disk with a dust temperature of $T_d = 20$ K, i.e.

\begin{eqnarray*}
M_d &=& \frac{F_{\nu} d^2}{\kappa_{\nu} B_{\nu}(T_d)}\\
\end{eqnarray*}
where $F_{\nu}$ is the integrated flux at 870 $\mu$m, $d$ is the estimated distance to Ophiuchus (137 pc), $\kappa_{\nu} = 0.03$ cm$^2$/g is the total opacity at 870 $\mu$m 
assuming a \cite{hi83} dust opacity and a 100:1 gas-to-dust ratio \citep{bo78}, and $B_{\nu}$ is the Planck function. 
An important caveat here is that this mass is calculation is only an estimate at best. Recent studies have suggested that the gas mass might be considerably lower than the often 
prescribed 100:1 ratio \citep{wi14}.

\begin{figure*}[htp]
\centering
\begin{tabular}{cc}
\includegraphics[scale=0.6]{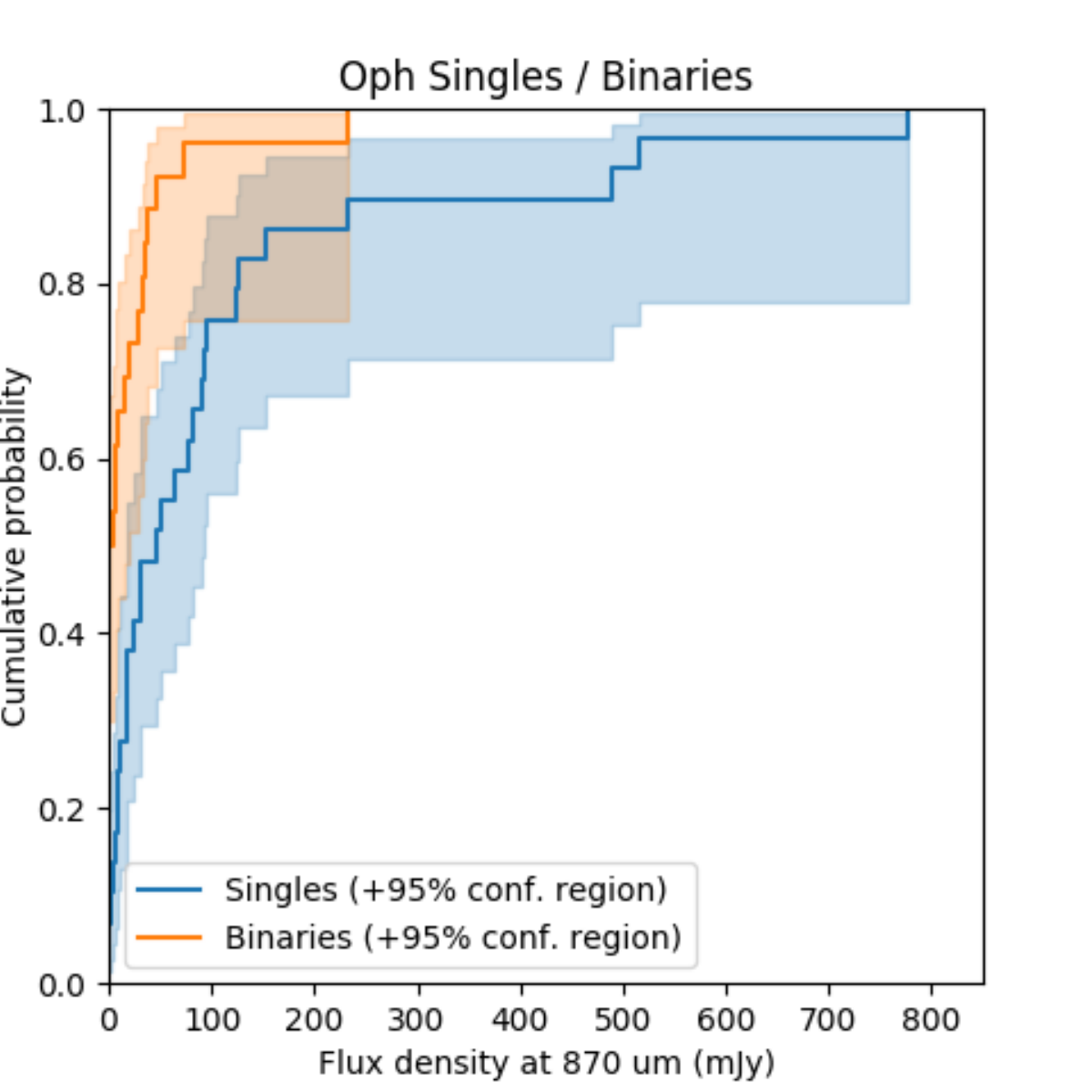}
\includegraphics[scale=0.6]{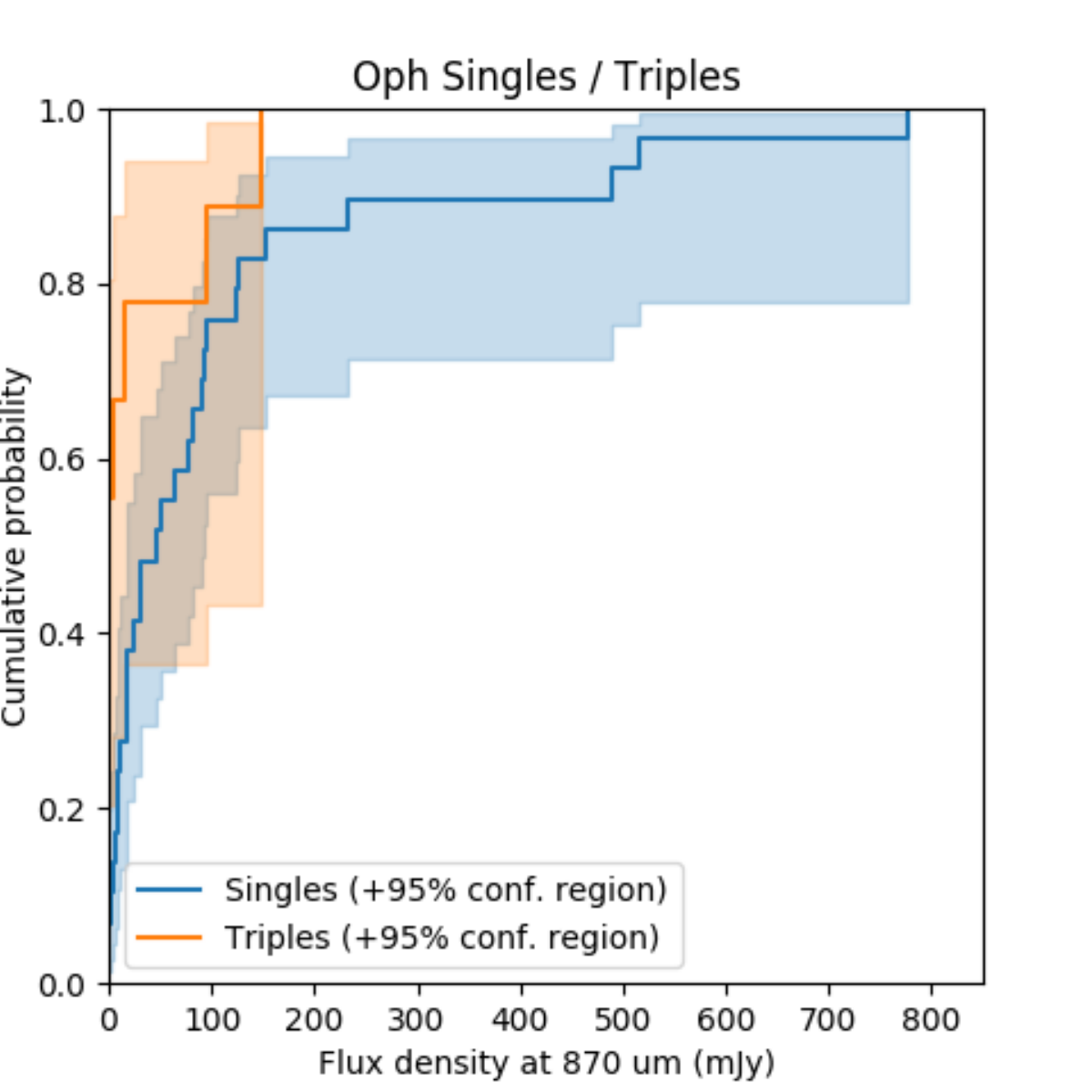}\\
\includegraphics[scale=0.6]{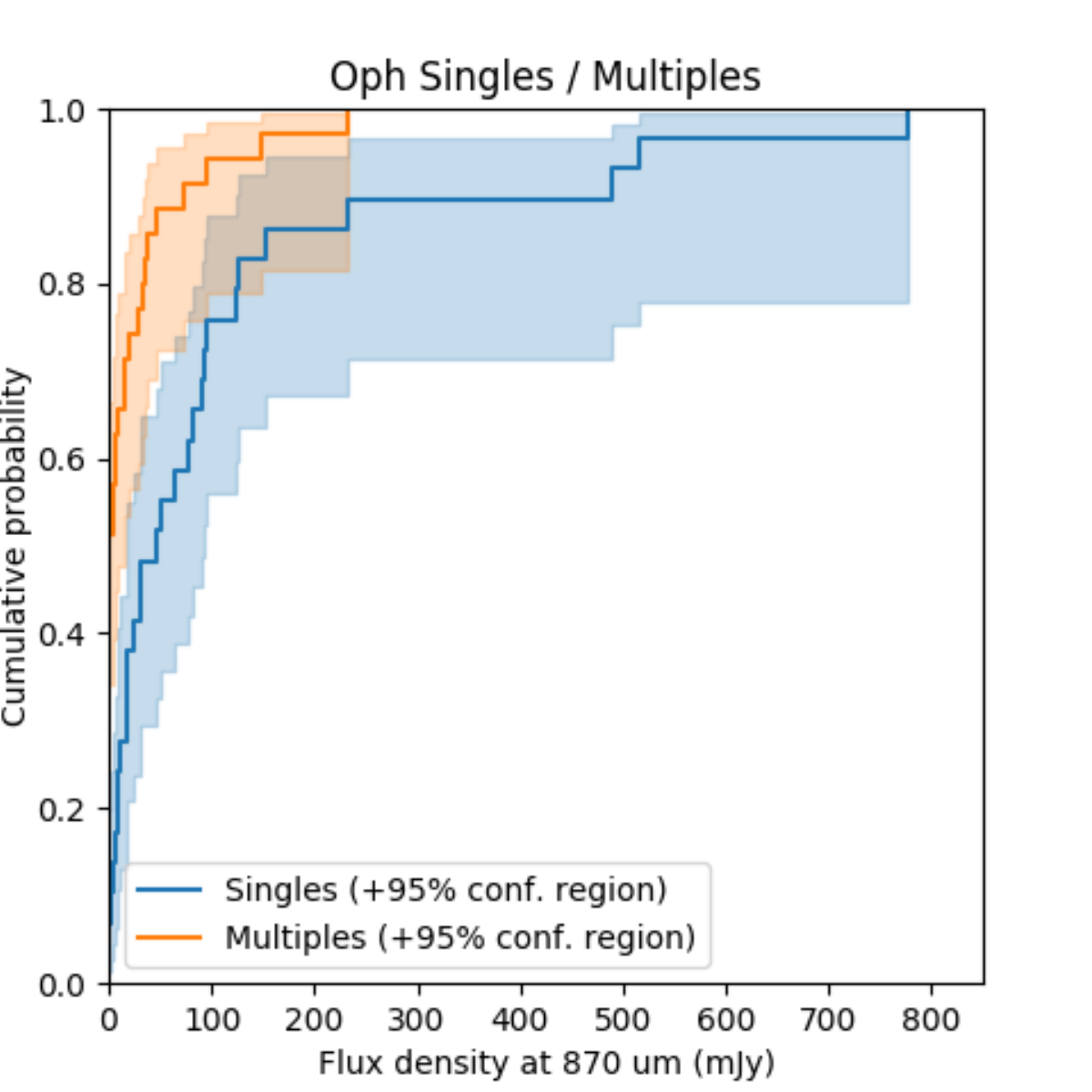}
\includegraphics[scale=0.6]{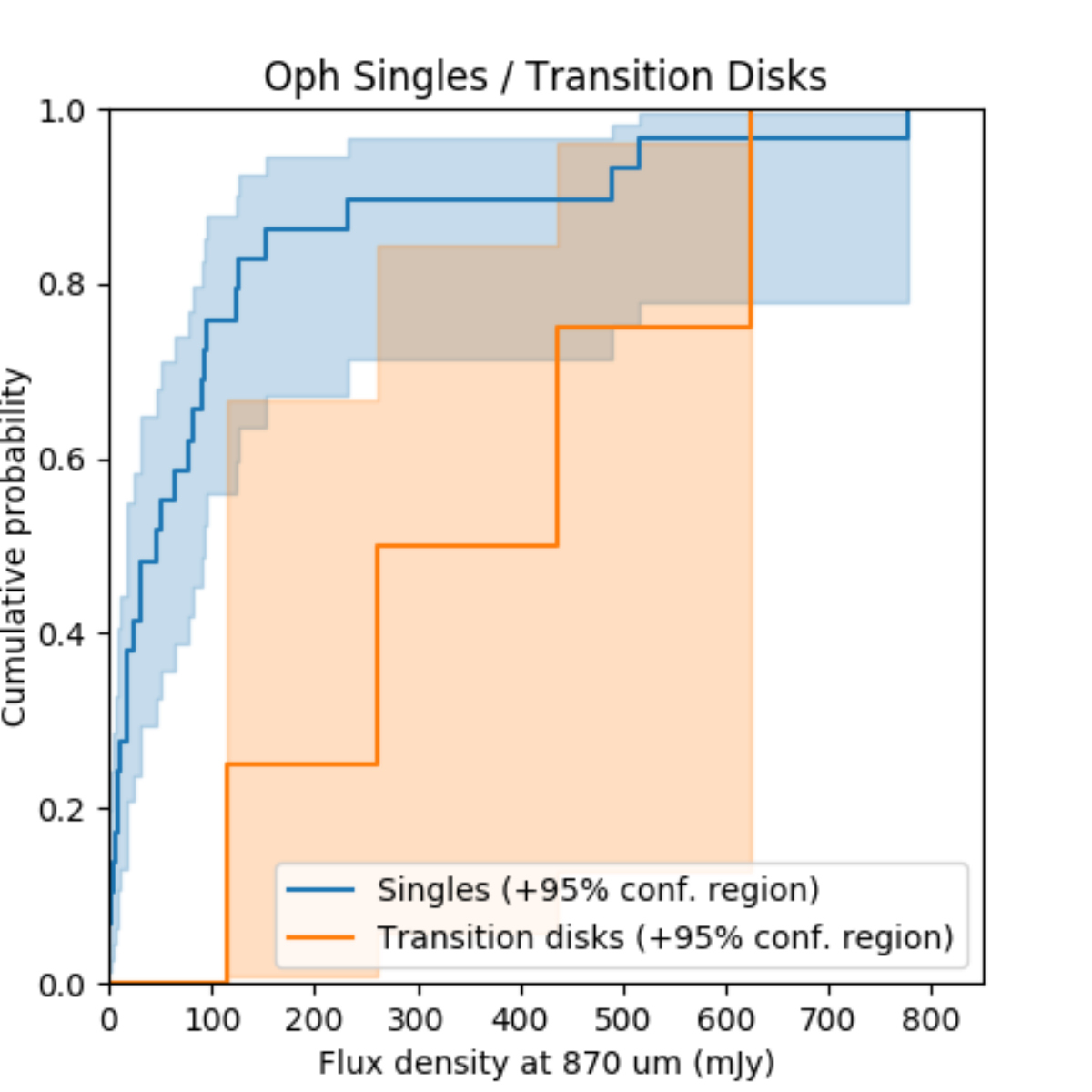}
\end{tabular}
\caption{CDF flux comparisons between the single population of protostars in $\rho$ Ophiuchus and the
other multiplicities. The shaded area is taking into account the upper limits of the various fluxes. Note that the transition disks are
only included in that category and not in the singles or triples.}
\label{fig:single_flux_cdfs}
\end{figure*}

\begin{figure*}[htp]
\centering
\begin{tabular}{cc}
\includegraphics[scale=0.6]{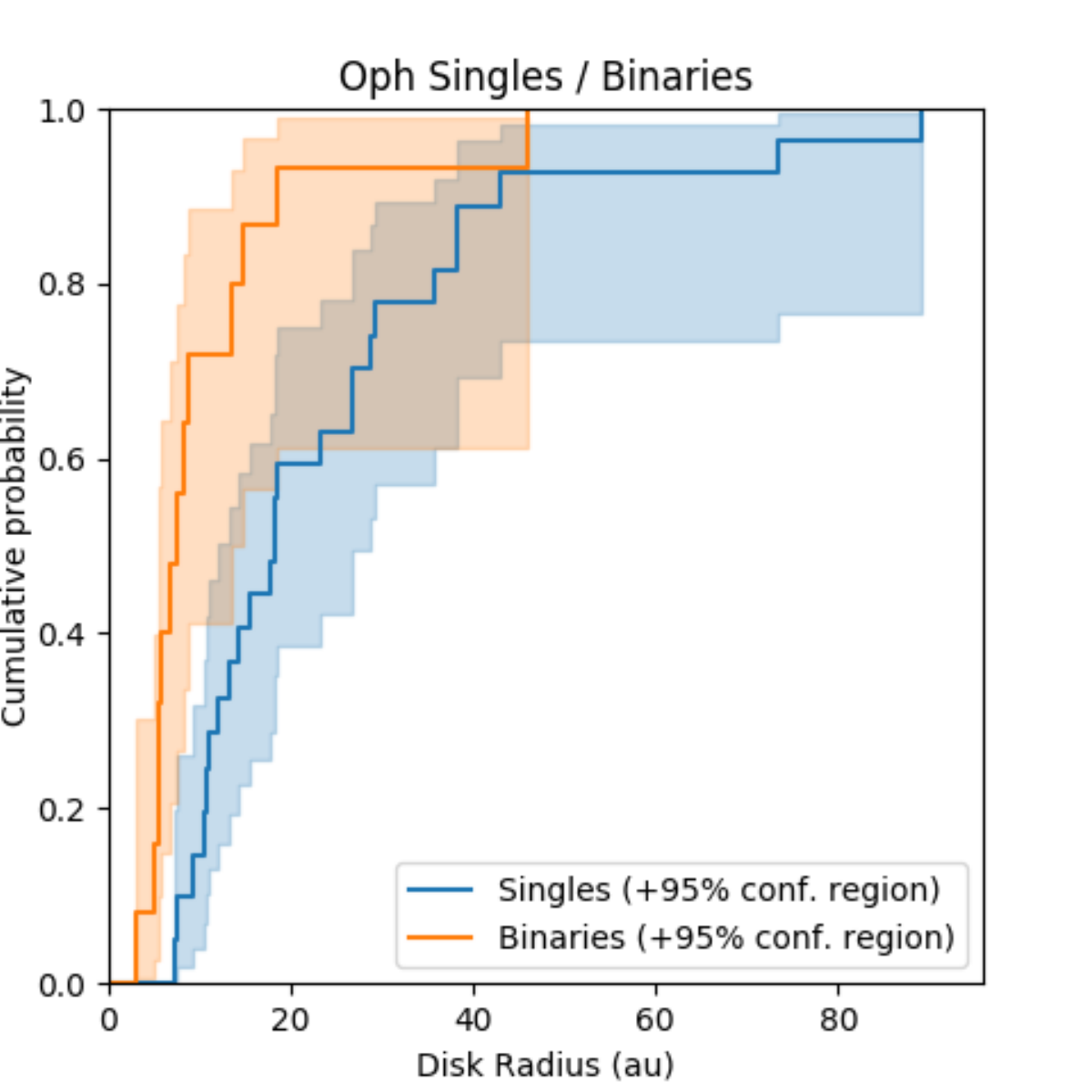}
\includegraphics[scale=0.6]{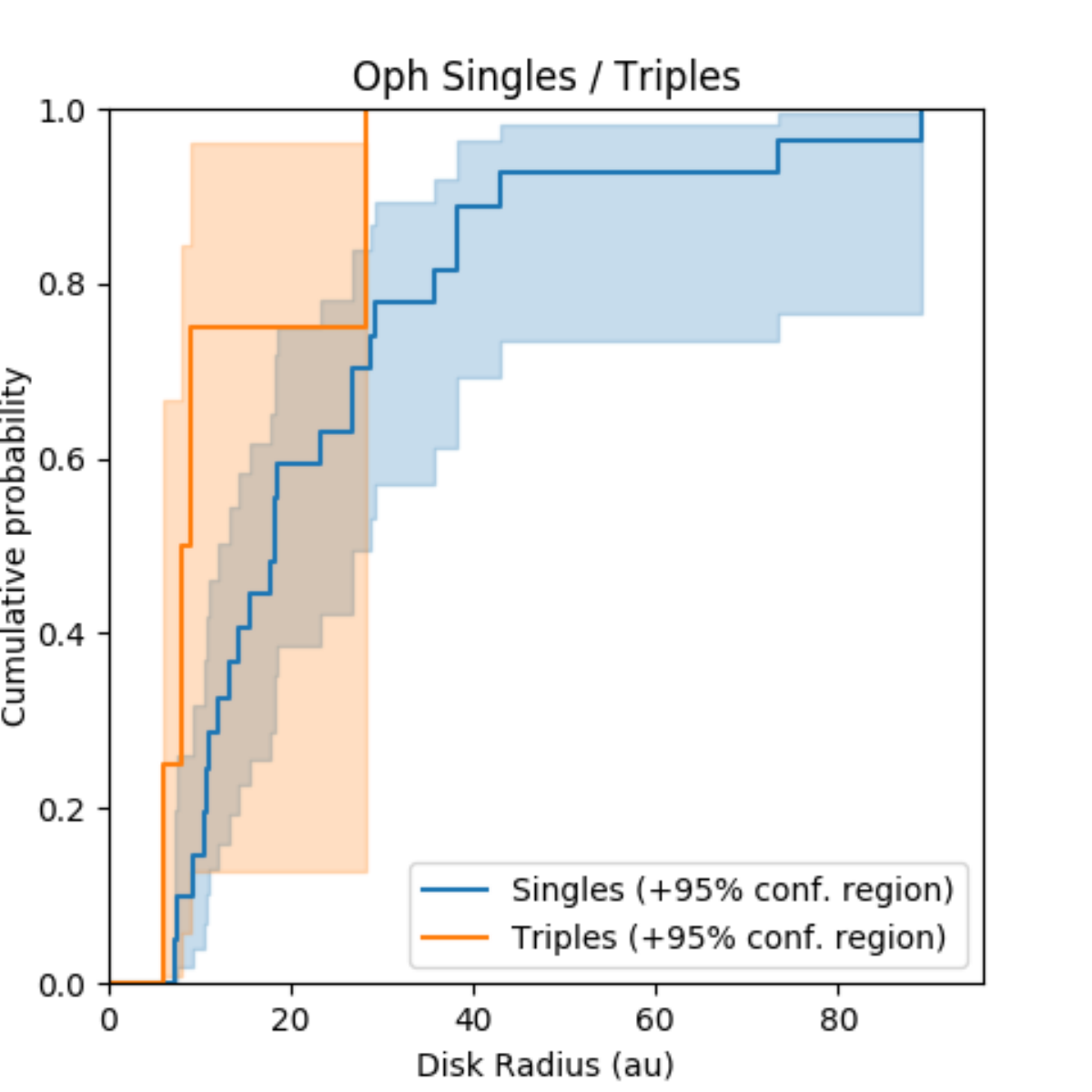}\\
\includegraphics[scale=0.6]{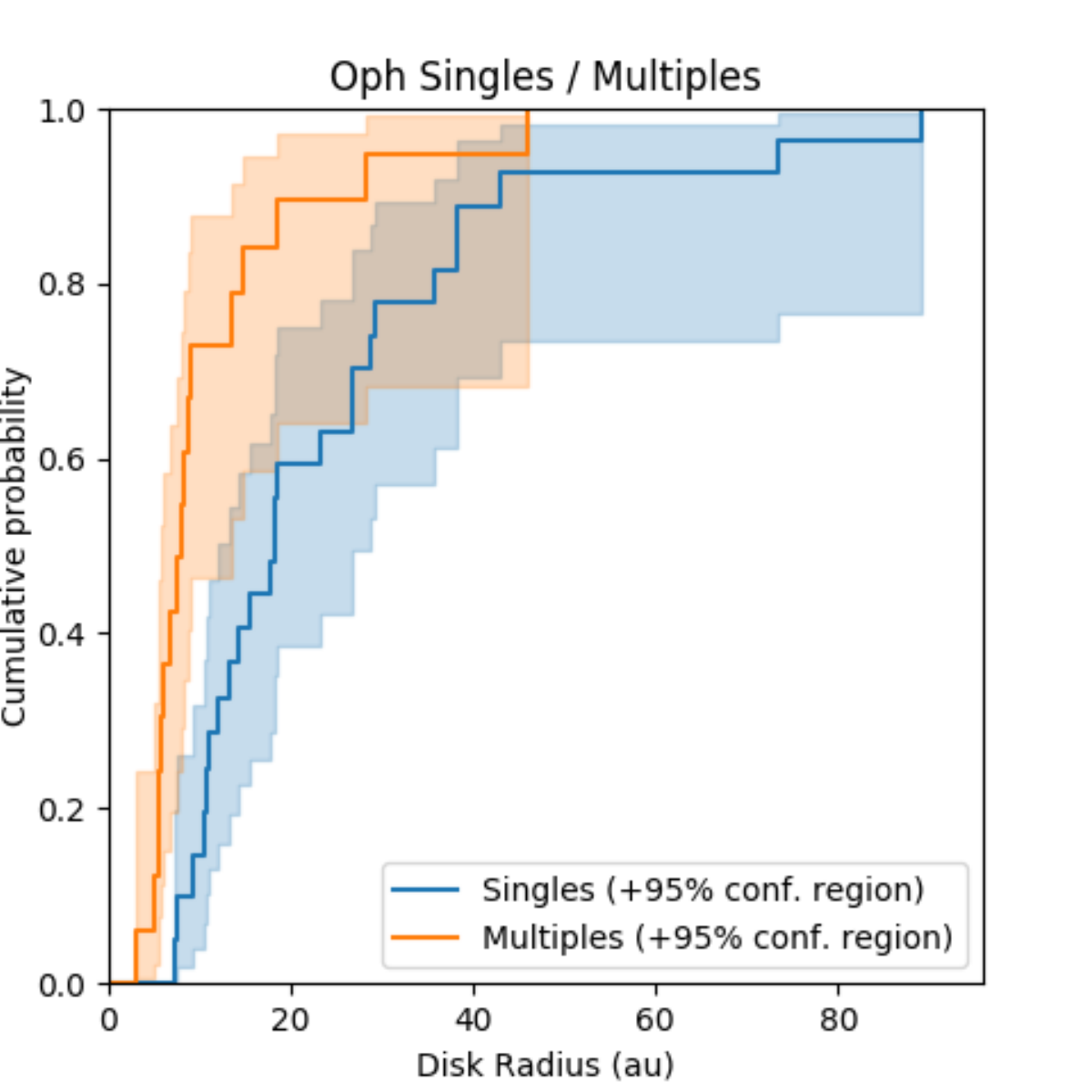}
\includegraphics[scale=0.6]{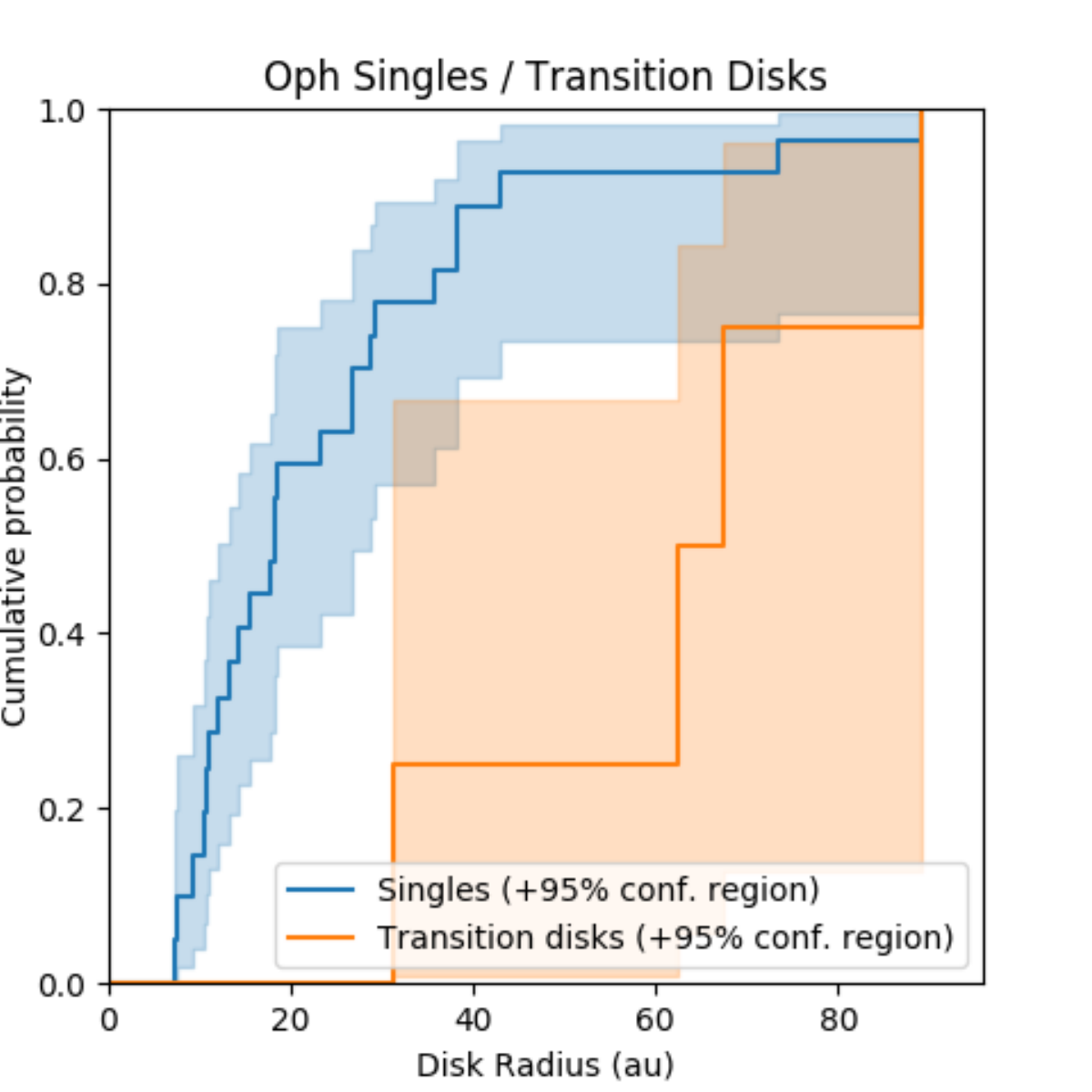}
\end{tabular}
\caption{Radius CDF comparisons between the single population of protostars in $\rho$ Ophiuchus and the 
other multiplicities. The shaded area is taking into account the upper limits of the radii. Note that the transition disks are
only included in that category and not in the singles or triples.}
\label{fig:single_radii_cdfs}
\end{figure*}

\begin{figure*}[htp]
\centering
\begin{tabular}{cc}
\includegraphics[scale=0.6]{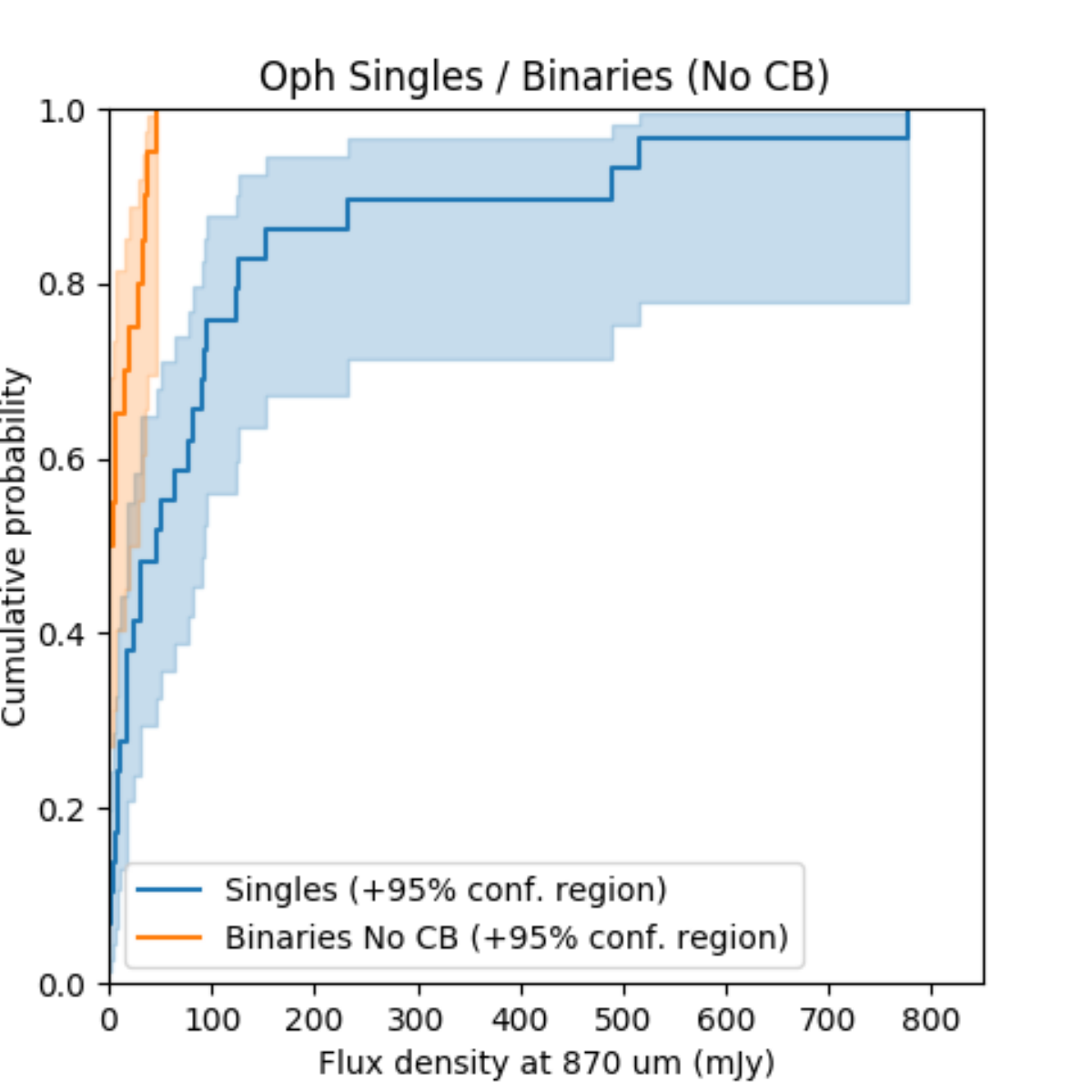}
\includegraphics[scale=0.6]{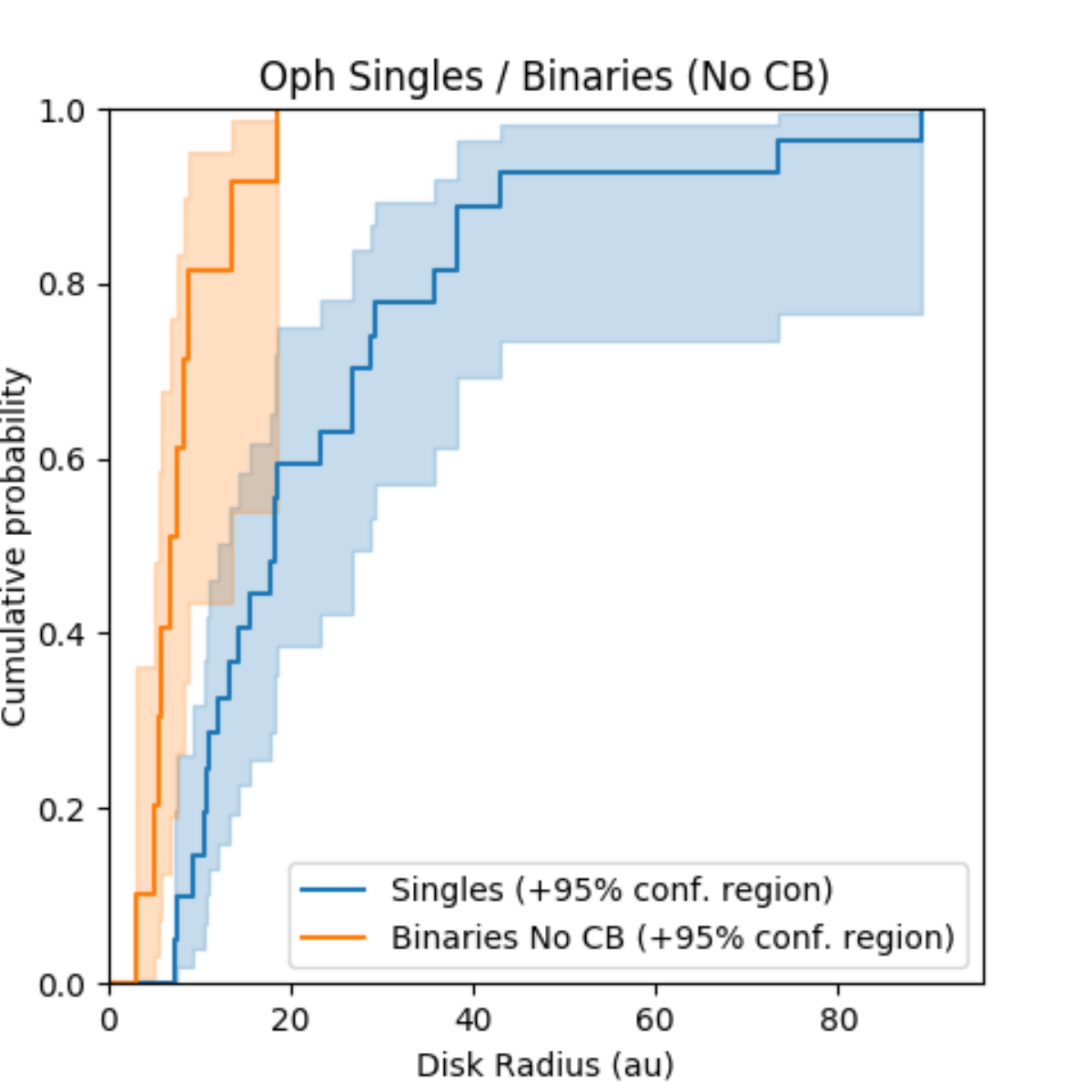}
\end{tabular}
\caption{CDF comparing the single and binary population of $\rho$ Ophiuchus, this time with the circumbinary disks
taken out of the sample. Note how different the two populations are when the circumbinary disks are taken out of the binary
sample.}
\label{fig:singles_vs_bin_nocb}
\end{figure*}

\begin{figure*}[htp]
\centering
\begin{tabular}{cc}
\includegraphics[scale=0.6]{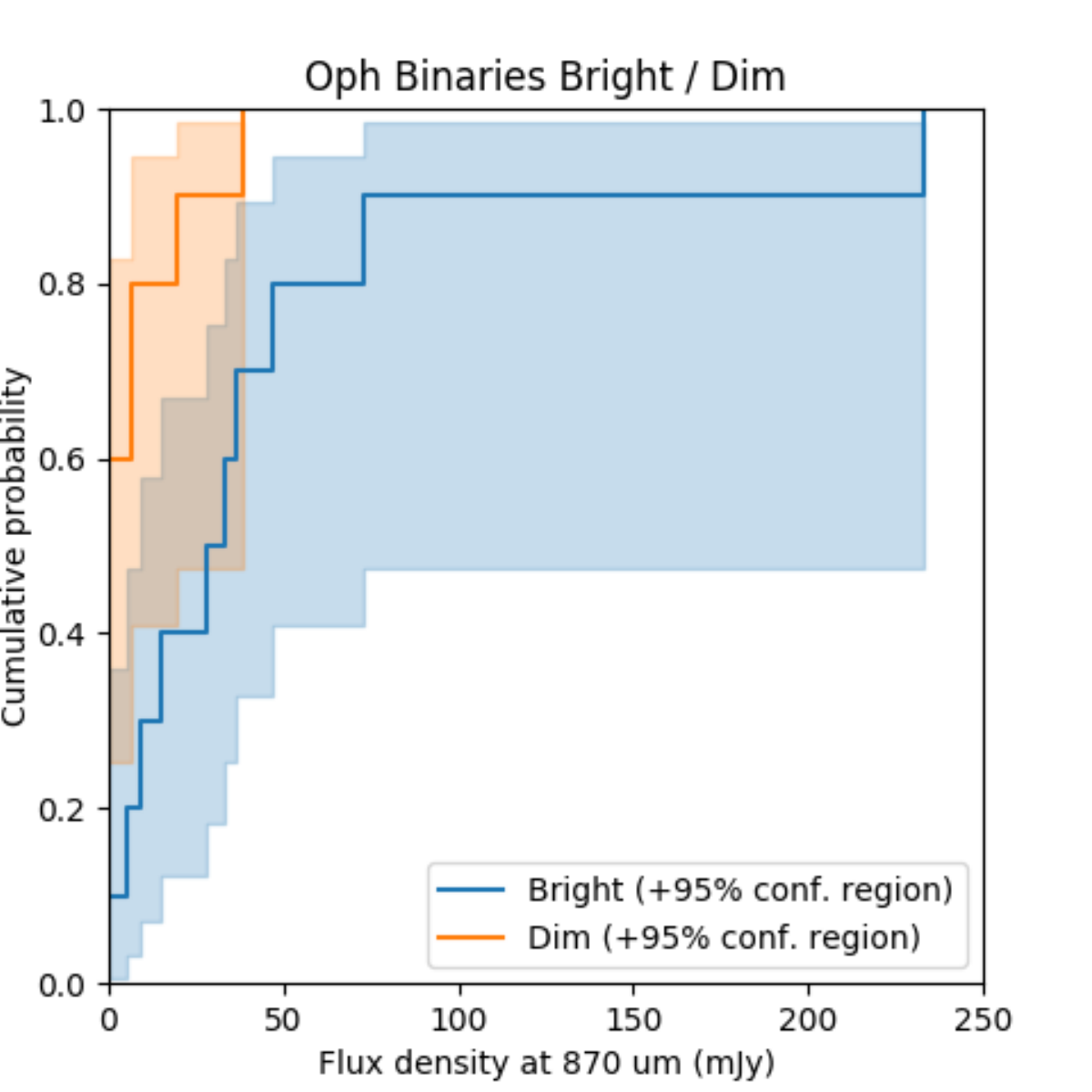}
\includegraphics[scale=0.6]{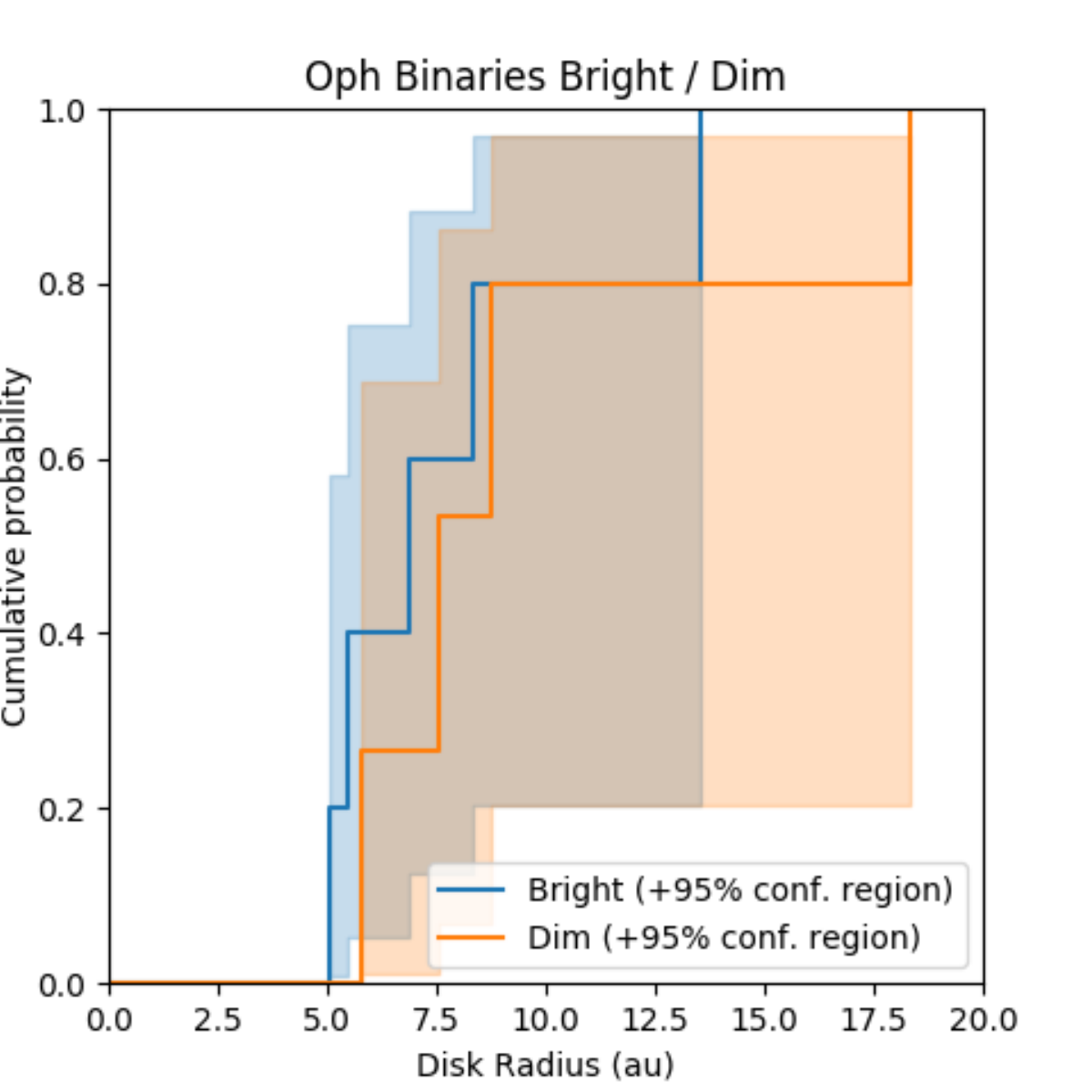}
\end{tabular}
\caption{CDF comparisons of both flux and radius for the brighter and dimmer component of the binary protostar
population in $\rho$ Ophiuchus.}
\label{fig:primary_vs_secondaries}
\end{figure*}

\begin{figure*}[htp]
\centering
\begin{tabular}{cc}
\includegraphics[scale=0.6]{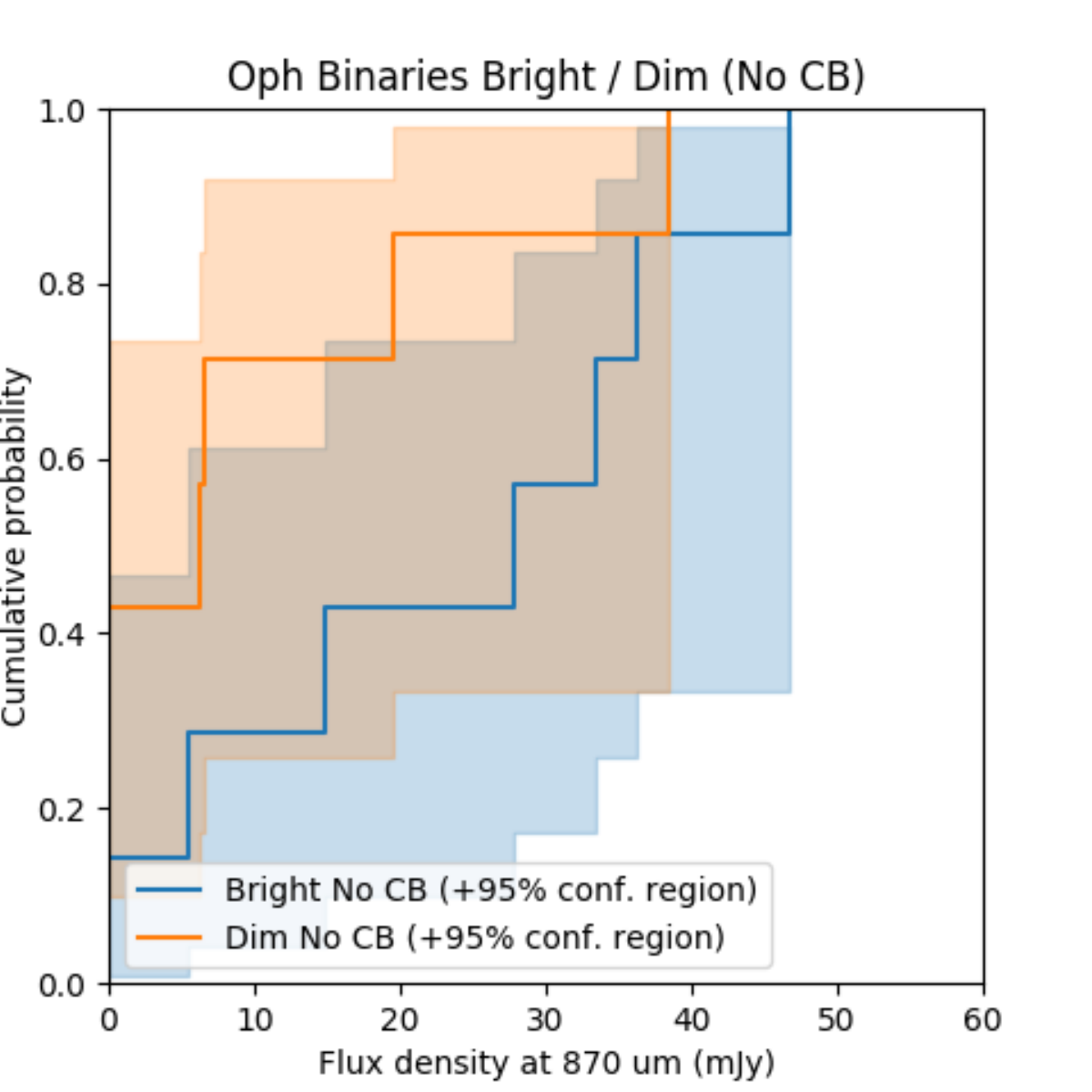}
\includegraphics[scale=0.6]{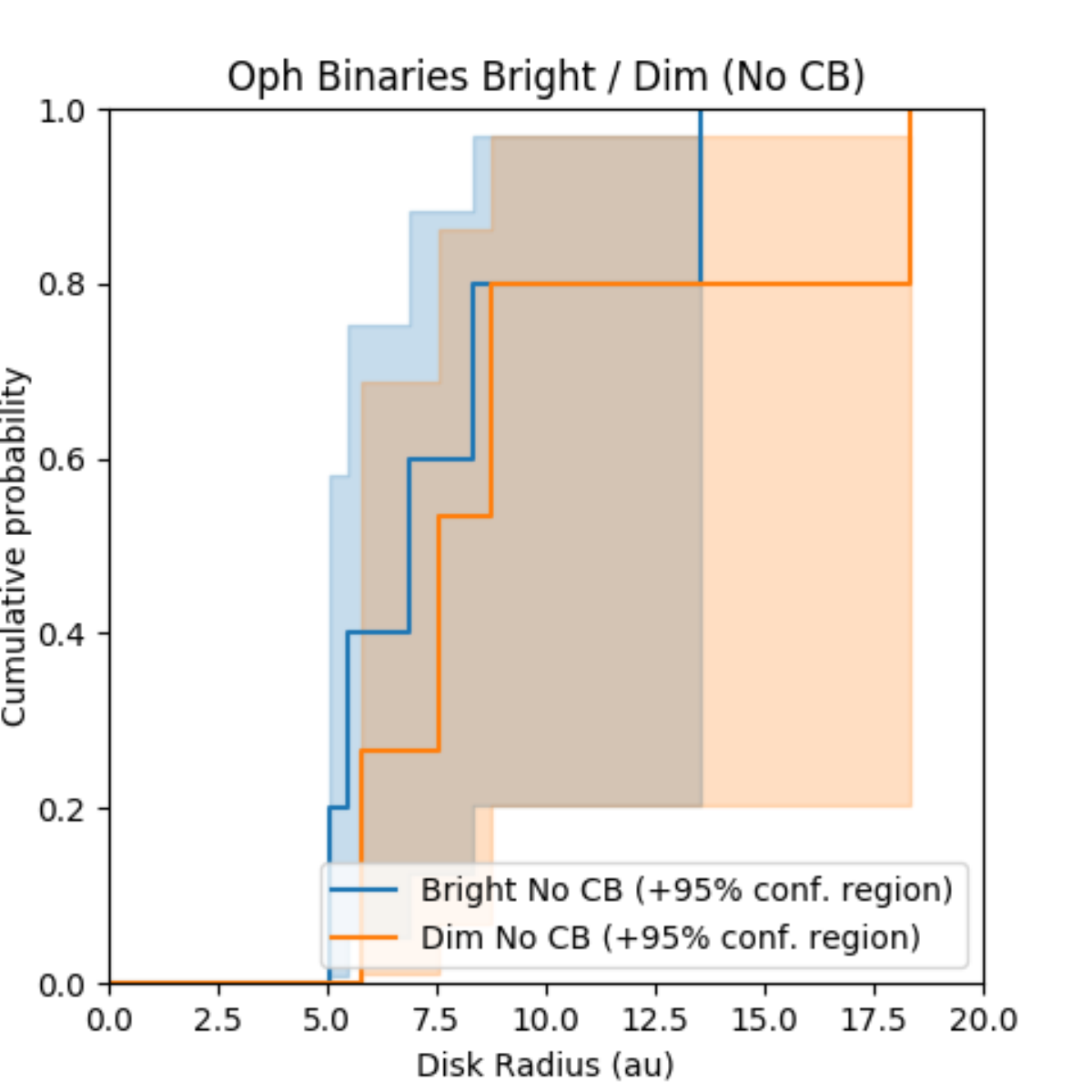}
\end{tabular}
\caption{CDF comparisons of both flux and radius for the brighter and dimmer component of the binary protostar
population in $\rho$ Ophiuchus, without Circumbinary disks.}
\label{fig:primary_vs_secondaries_nocb}
\end{figure*}

\begin{figure*}[htp]
\centering
\begin{tabular}{cc}
\includegraphics[scale=0.6]{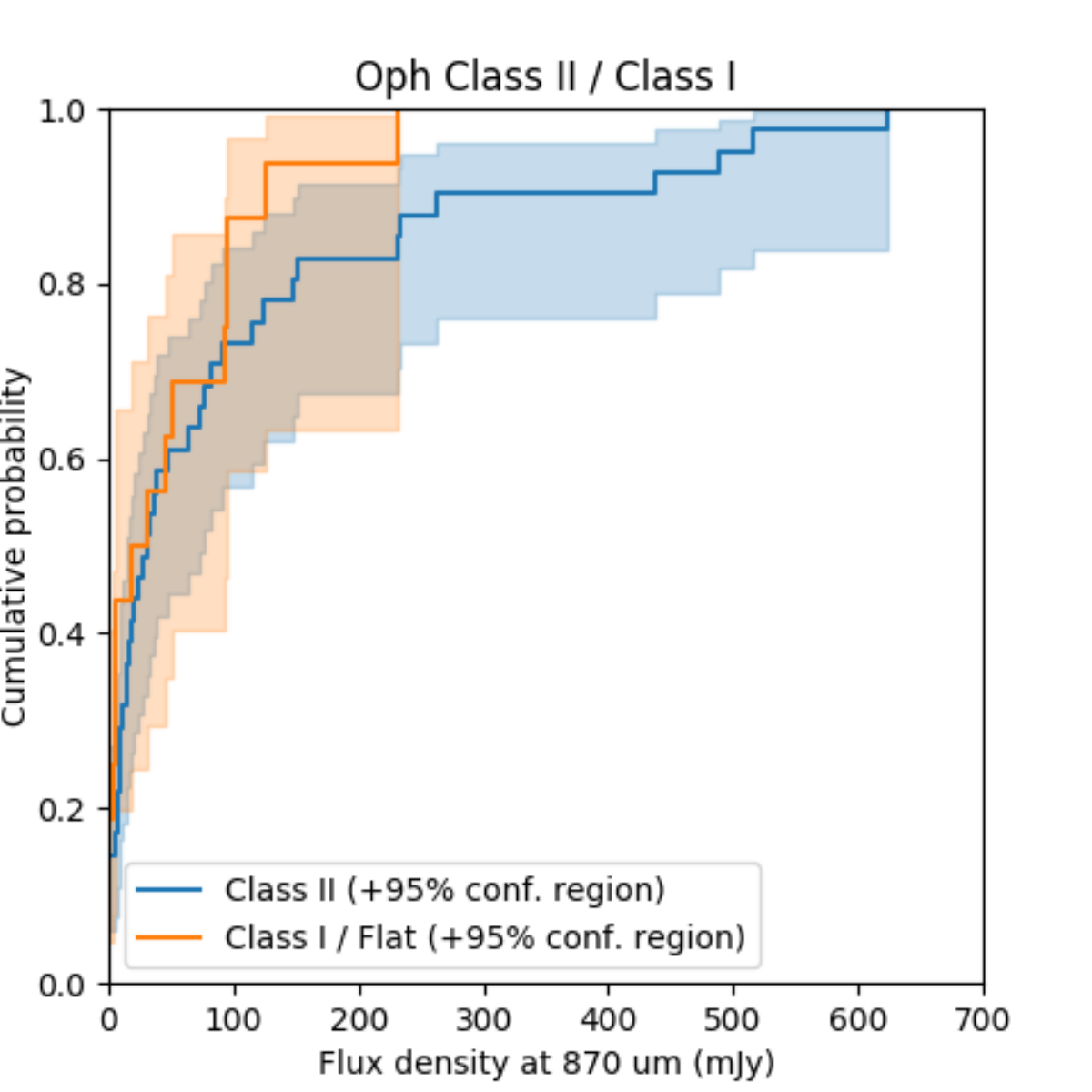}
\includegraphics[scale=0.6]{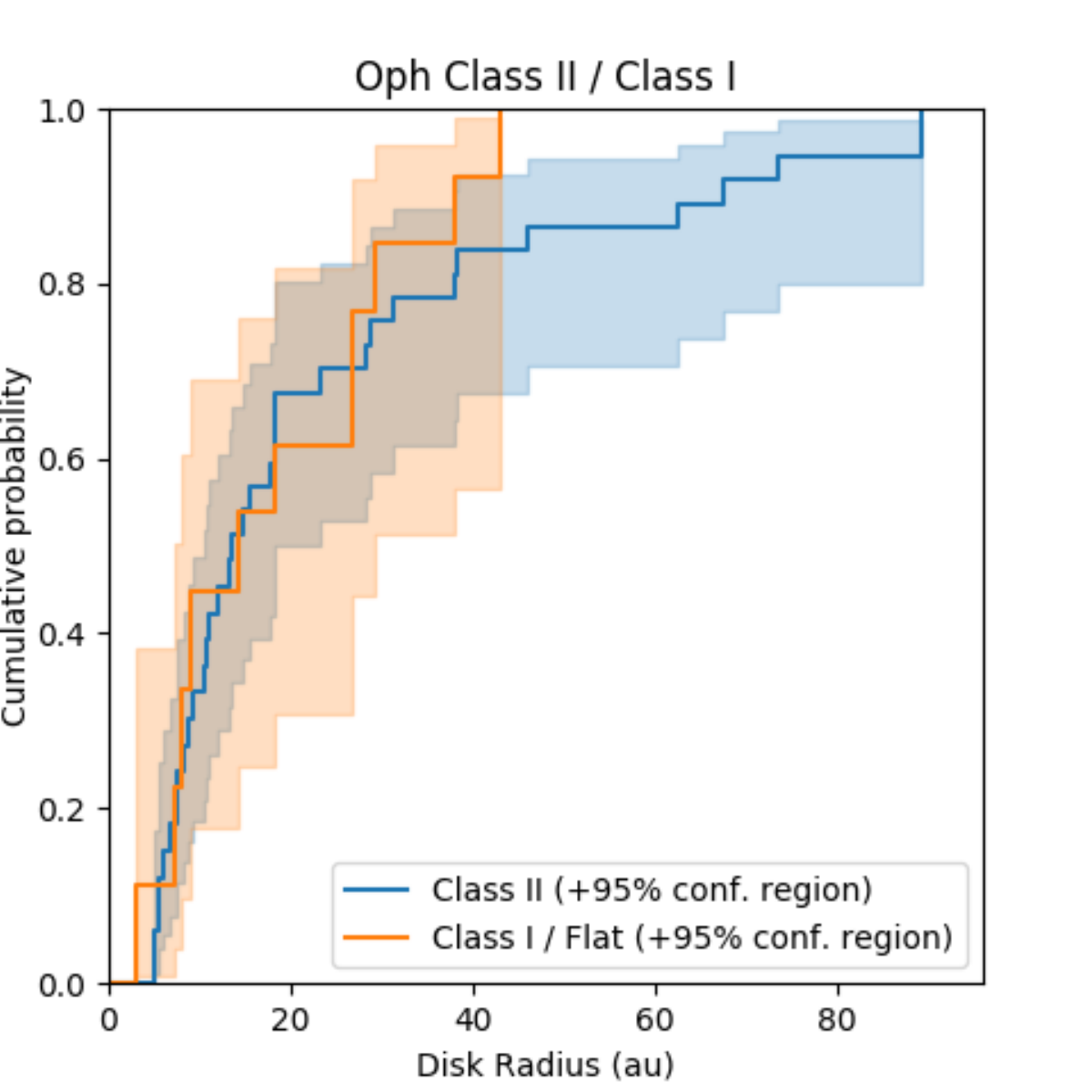}
\end{tabular}
\caption{CDF comparisons of both flux and radius for Class II sources and Class I/Flat sources
in $\rho$ Ophiuchus.}
\label{fig:cdfs_class1_vs_2}
\end{figure*}

To test for significant differences in flux and radius among the different sub-populations of our sample, we used the implementation of the Kaplan-Meier (KM) product estimator in the \texttt{lifelines} Python package \citep{dp17} 
\footnote{This package is available at \url{https://github.com/CamDavidsonPilon/lifelines/}} to estimate the Cumulative Distribution Functions (CDFs) for all of the sub-populations
(see Figures \ref{fig:single_flux_cdfs} - \ref{fig:tauclass}). The KM estimator is akin to an empirical cumulative probability distribution, but it has the advantage of being able to account for the non-detections in our sample
by incorporating $\sigma$ upper limits when appropriate. For all the distributions we compute for fluxes, upper limits are incorporated. However, for the radii KM estimators, we only incorporate detections, as the radius of a non-detected object is ill-defined. Confidence intervals for each bin in the KM estimator are computed 
using Kalbfleisch and Prentice's modification of the result of 
\cite{gw26} (see p. 18 of \citealt{kp02} for details).

After the KM estimators are computed for each sub-population, we use the non-parametric log-rank test to determine whether or not it is likely that the two cumulative distributions in flux or radius are different for the pairwise combinations of sub-populations. 
Figure \ref{fig:single_flux_cdfs} shows the CDFs comparing the flux of the single sources in our 
survey with the fluxes of the other populations (binary, triples, 
multiples, and transition disks). Perhaps most striking of these 
comparisons, is that of the binary population. The binary sources in this
survey show systematically lower flux values than the isolated population. 

In Table 3 we report the $p$-values of the different comparisons, as well as
the median flux and radius of the different populations. The $p$-value
 represents the probability that, given our data, the two populations compared are
drawn from a single distribution. Thus, the higher our $p$-value, the more likely
that this is the case; conversely, the lower the $p$-value, the less
likely it is for the two populations to be from the same distribution.  We define two populations to have a significant difference if the log-rank on their respective KM estimators yields $p\lesssim0.05$.

When comparing the binaries with the singles, we find that both the flux and radius $p$-values 
show a suggestive trend with both $p$-values $<0.1$ ($p_{flux}=0.06946$ and $p_{radius}=0.01766$).
Our binary sample includes three circumbinary disks (disks that encompass both components of the binary system), that we find 
to be quite bright compared to the rest of the binary sample. Since \citet{ha12} also found this to be true in Taurus, we looked at the
same comparison without these sources. We find that without the circumbinary disks, we get $p_{flux}=0.00876$ and 
$p_{radius}=0.00075$, which is lower than our cutoff.
Figures \ref{fig:single_flux_cdfs} and \ref{fig:single_radii_cdfs} show the CDF comparing the two populations in the top left panel
for flux and radius, respectively, and Figure \ref{fig:singles_vs_bin_nocb} shows the
same plot excluding circumbinary disks. Each binary component was counted as one source,
and in the case of a non-detection, the 3$\sigma$ value of the map was used as an upper limit for fluxes. 
All known, non-spectroscopic, (i.e., Oph 6, 12, 32, and 36) binaries in our sample are resolved, therefore blending of component fluxes is not an issue in our sample.
It can be visually seen in these plots (see top left panel of both Figure \ref{fig:single_flux_cdfs} and Figure \ref{fig:single_radii_cdfs}, and Figure \ref{fig:singles_vs_bin_nocb}), that there
is hardly any overlap between the isolated YSOs and the binaries. The binary components are systematically
dimmer as well as smaller, than their isolated counterparts. In Figure \ref{fig:primary_vs_secondaries}, we show the comparison between the components of  the binary YSOs. The brighter component
has a median flux value of 27.74 mJy, while the dimmer component is at 6.45 mJy. This a factor of 5 different, although we note that the large uncertainty in each individual bin of the KM estimator makes any observed difference between the populations not significant. 
The difference in the median radius for either component is $<$ 2 au, meaning that there is not a discernible
difference in the sizes of the two. 

Across star forming regions, the inner disk fraction for single stars and for wide binaries (i.e., binaries with projected separation $>$ 40 au) is comparable, with $\sim$ 50\% of 
these systems harboring enough material to make them Class II objects. On the other hand, tighter systems  ($<$ 40 au) are preferentially less likely to have evidence of an 
infrared excess, with only $\sim$ 20\% of those systems harboring enough material to make them Class II objects \citep{ci09}. Folding these data into our analysis of the 
millimeter emission for singles vs. binaries would most likely make the difference between the two much starker.

The triple systems in our sample are slightly more complicated than the binaries. Two of the three systems (ROph11 and ROph31) are treated
in the same way as the binaries, where we use the 3$\sigma$ value for the non-detections. The third system, ROph23, is also treated in 
this way, however this system has a circumbinary disk. Since this cannot be 
divided into two different systems, we count this as one source and use the 3$\sigma$ value of the map as the upper limit 
twice. When comparing these with the singles, we find $p_{flux}=0.73140$ and $p_{radius}=0.03613$. A caveat to keep in mind
when looking at the triple systems in this sample, is that we did not detect all three sources in any of the systems. These systems
consist of a tight pair that will resemble binary systems, with a single star further away. In ROph 11 and in ROph 31, the distance of the third component from the tight pair is much larger than the separation of the tight pair itself. Therefore, the disk associated with the distant object more closely resembles a disk from a single source. 
This is likely the case in ROph 23 as well, though the orbit superimposes the distant companion onto the circumbinary disk. 

The transition disks we used in comparing with the isolated sources were ones that show a depletion of 
millimeter emission in their inner cavity
in this dataset, not necessarily those listed in Table 2. 
Visually, our transition disk population (see Figure \ref{fig:trans}) seems to be the most unique in both flux and size. 
It was somewhat surprising
that the fluxes of these disks did not show $p<0.05$ when compared to the singles ($p_{flux}=0.10204$ and $p_{radius}=0.04363$).
We only have 4 transition disks in our sample, so the small numbers may contribute to the higher $p$-values. The median flux
for this population is a factor of 5 brighter than any other population and the median radius
is 3.5 times as large as the isolated population, suggesting
that the transition disks come from a different distribution. We did use the two different populations of Table 2 to see
if there was anything statistically different between sources that either have sub-/millimeter cavities or infrared colors indicating
they are transition disks, and those that do not. We find $p_{flux}=0.12846$ and $p_{radius}=0.09715$. 

Figure \ref{fig:cdfs_class1_vs_2} shows the CDF plots for the different classifications of YSOs. Our sample consisted of mostly Class II
YSOs, followed by Flat and then Class I objects. Since the Flat sources are thought to be on average less evolved than the the Class II 
sources, we combined these with the Class I sources to more easily compare the two. The less evolved 
population shows a higher ($\sim$ 13 mJy) median flux, while also having a slightly lower ($\sim$ 1 au)
median radius. This is as expected, since as the YSO evolves into a Class II object, its peak energy output
moves to shorter wavelengths \citep{la87} and, as the nascent material from the envelope falls in,
the disk surrounding the protostar will grow \citep{du14}.

We report the detection of a $\sim 1600$ au millimeter-wave companion to ROph34, L1689-IRS 7. The system L1689 IRS 7 has only sparsely been surveyed for companions. It was 
included in the Ratzka et al. (2005) survey area, but the source was determined to be single. The separation regime that the Ratzka survey was sensitive to ranged from 0.1 to 6.4 
arcseconds, and the companion that we report is located outside of 7 arcseconds.  The companion can be seen in 2MASS. The $JHK_s$ magnitudes of the northern component are 
uniformly $\sim$ 2 mag lower than the corresponding magnitudes for the southern component $K_s \sim 8.5$ for the primary and $10.5$ for the secondary. Since the colors are the 
same, it is likely that the companion is also a class II low mass star that is a bona fide member of the Oph complex. From the K band contrast, we estimate a stellar mass ratio of 0.1-0.3 
based on Seiss (2001) models for a 1 Myr old object (roughly consistent with the fact that the source is a Class II object).

\begin{table}[h!]
\begin{center}
\caption{Statistics of different populations.}
\vspace{0.2cm}
\begin{tabular}{|r|r|r|}
\hline
Comparison & Flux $p$-value & Radius $p$-value\\ \hline
Singles \& Binaries & 0.06946 & 0.01766 \\
Singles \& Binaries (no circumbinary disk) & 0.00876 & 0.00075 \\
Singles \& Triples & 0.73140 & 0.03613 \\
Singles \& Multiples & 0.11562 & 0.00368 \\
Singles \& Transition Disks & 0.10204 & 0.04363 \\ 
Binary Components & 0.21271 & 0.15456 \\
Binary Components (no circumbinary disk) & 0.53125 & 0.15456 \\
Class II \& Class I/Flat & 0.24451 & 0.79303 \\ \hline \hline
Population & Median Flux (mJy) & Median Radius (au)\\ \hline
Singles & 46.3 & 17.9\\
Binaries & 27.74 & 7.1 \\
Binaries (no circumbinary disk) & 19.6 & 6.85 \\
Bright Binary component & 27.74 & 6.45 \\
Bright Binary (no circumbinary disk) & 21.29 & 6.17 \\
Dim Binary component & 6.17 & 7.54 \\
Dim Binary (no circumbinary disk) & 6.45 & 7.54 \\
Triples & 15.41 & 8.08\\
Multiples & 19.6 & 7.8\\
Transition Disks & 262 & 62.34 \\ 
Class I/Flat Sources & 30.55 & 12.6 \\
Class II Sources & 18.73 & 13.426\\
\hline
\end{tabular}  
\end{center}
\scriptsize Comparison of the various $p$-values obtained from each CDF. Note that 'multiples' represents a combination of both binary and triple systems. We used a distance of 137 pc to compute the
radius
\end{table}

\begin{longrotatetable}
\begin{deluxetable}{clclccccll}
\setlength{\tabcolsep}{5pt}
\tabletypesize{\scriptsize}
\rotate
\tablecaption{Protoplanetary Disks in $\rho$ Ophiuchus Molecular Cloud.}
\tablewidth{0pt}
\tablehead{
\colhead{Index} & \colhead{Object} & \colhead{Class} & \colhead{Derived position} & \colhead{Disk size} & \colhead{Disk PA} & \colhead{Peak Flux} & \colhead{Integrated Flux}  & \colhead{$M_{\mathrm{disk}}$}\\
 \colhead{} & \colhead{} & \colhead{} & \colhead{(J2000)} & \colhead{(\arcsec)} & \colhead{($^{\circ}$)} & \colhead{(mJy beam$^{-1}$)} & \colhead{(mJy)} & \colhead{($M_{\mathrm{Jupiter}}$)}}
\startdata
1 & 2MASS J16213192-2301403 & II &16:21:31.923 -23:01:40.761& 0.193 $\pm{0.015}$ $\times$ 0.07 $\pm{0.032}$ & 164 $\pm{6.6}$ & 6.2 $\pm{0.21}$ & 9.63 $\pm{0.49}$  & 0.59 \\
2$^{a,*}$ & V935 Sco & II &16:22:18.523 -23:21:48.549& 0.215 $\pm{0.011}$ $\times$ 0.125 $\pm{0.009}$ & 80.5 $\pm{4.7}$ & 41.98 $\pm{0.89}$ & 72.9 $\pm{2.3}$ &  4.43 \\
3$^{c}$ &	IRAS 16201-2410 & II &16:23:09.219 -24:17:05.364& 0.456 $\pm{0.043}$ $\times$ 0.283 $\pm{0.029}$ & 81.4 $\pm{8.2}$ & 26 $\pm{1.8}$ & 114.3 $\pm{9.7}$ & 6.96 \\
4 & 2MASS J16233609-2402209 & II &16:23:36.113 -24:02:21.227& 0.160 $\pm{0.014}$ $\times$ 0.072 $\pm{0.016}$ & 6.7 $\pm{6.5}$ & 5.27 $\pm{0.15}$ & 7.12 $\pm{0.31}$ & 0.43 \\
5$^{a}$a & 	WSB 19A  & II &16:25:02.119 -24:59:32.798& 0.198 $\pm{0.011}$ $\times$ 0.188 $\pm{0.013}$ & 145 $\pm{60}$ & 14.11 $\pm{0.27}$ & 27.74 $\pm{0.77}$ & 1.69 \\
5$^{a}$b & WSB 19B & II &16:25:02.011 -24:59:33.004& 0.127 $\pm{0.017}$ $\times$ 0.107 $\pm{0.019}$ & 106 $\pm{39}$ & 14.45 $\pm{0.33}$ & 19.60 $\pm{0.7}$ & 1.20 \\
6 & DoAr 21 & II &16:26:03.300 -24:23:36.000& -- $\times$ --& ... & ... & $<$ 0.89 & $<$ 0.05 \\
7$^{a}$a & GSS 31a & II &16:26:23.362 -24:20:59.997& 0.100 $\pm{0.004}$ $\times$ 0.060 $\pm{0.007}$ & 169 $\pm{5}$ & 39.34 $\pm{0.21}$ & 46.75 $\pm{0.4}$ & 2.85 \\
7$^{a}$b & GSS 31b & II &16:26:23.432 -24:21:01.749& 0.071 $\pm{0.004}$ $\times$ 0.051 $\pm{0.008}$ & 147 $\pm{15}$ & 34.67 $\pm{0.21}$ & 38.36 $\pm{0.38}$ & 2.33 \\
8 & DoAr 25 & II &16:26:23.680 -24:43:14.303& 1.071 $\pm{0.038}$ $\times$ 0.487 $\pm{0.015}$ & 110 $\pm{1.4}$ & 38 $\pm{1.2}$ & 515 $\pm{18}$ & 31.33 \\
9$^{d}$ &Elias 24 & II &16:26:24.078 -24:16:13.855& 0.558 $\pm{0.049}$ $\times$ 0.474 $\pm{0.042}$ & 45 $\pm{81}$ & 63 $\pm{4.5}$ & 489 $\pm{39}$ &  29.77 \\
10 & GY 33 & II &16:26:27.540 -24:41:53.882& 0.337 $\pm{0.013}$ $\times$ 0.097 $\pm{0.014}$ & 160.5 $\pm{1.7}$ & 10.22 $\pm{0.27}$ & 23.45 $\pm{0.84}$ & 1.43 \\
11$^{t}$ & WSB 38 Aa & II &16:26:46.471 -24:12:00.39&  ... & ...  & $<$0.8 & $<$0.8 & $<$0.04 \\
11$^{t}$ & WSB 38 Ab & II &16:26:46.474 -24:12:00.30& ... & ... &$<$0.8 & $<$0.8 & $<$0.04  \\
11$^{t}$ & WSB 38B & II &16:26:46.427 -24:12:00.443& 0.086 $\pm{0.013}$ $\times$ 0.033 $\pm{0.020}$ & 108 $\pm{14}$ & 13.82 $\pm{0.2}$ & 15.41 $\pm{0.38}$ & 0.94 \\
12$^{a,*?}$ & WSB 40 & II &16:26:48.651 -23:56:34.589& 0.082 $\pm{0.012}$ $\times$ 0.056 $\pm{0.018}$ & 167 $\pm{32}$ & 7.78 $\pm{0.13}$ & 8.83 $\pm{0.24}$ & 0.53\\
13$^{b}$ & SR 24aa & II &16:26:58.438 -24:45:32.24& ... & ... &  $<$ 1.2 & $<$ 1.2& $<$ 0.07\\
13$^{b}$$ $ & SR 24ab & II &16:26:58.453 -24:45:32.21& ... & ... & $<$ 1.2 & $<$1.2 & $<$ 0.07 \\
13$^{b,c}$ & SR 24b & II &16:26:58.504 -24:45:37.220& 0.984 $\pm{0.118}$ $\times$ 0.563 $\pm{0.070}$ & 22.5 $\pm{8.3}$ & 42.9 $\pm{4.7}$ & 624 $\pm{73}$  & 37.96 \\
14 & GY 211 & II &16:27:09.096 -24:34:08.708& 0.265 $\pm{0.005}$ $\times$ 0.127 $\pm{0.004}$ & 33.1 $\pm{1.2}$ & 45.79 $\pm{0.43}$ & 91.2 $\pm{1.2}$ & 5.55\\
15 & GY 224  & F &16:27:11.168 -24:40:47.100& 0.428 $\pm{0.009}$ $\times$ 0.148 $\pm{0.005}$ & 92.23 $\pm{0.79}$ & 42.13 $\pm{0.62}$ & 126.2 $\pm{2.4}$ & 7.67 \\
16 & GY 235 & F &16:27:13.813 -24:43:32.053& 0.208 $\pm{0.010}$ $\times$ 0.172 $\pm{0.010}$ & 177 $\pm{13}$ & 26.18 $\pm{0.41}$ & 51 $\pm{1.1}$ &  3.11 \\
17 & GY 284 & F &16:27:30.841 -24:24:56.528&  -- $\times$ -- & ...& 2.78 $\pm{0.11}$ & 2.79 $\pm{0.19}$ & 0.17 \\
18 & YLW 47 & II &16:27:38.314 -24:36:58.997& 0.155 $\pm{0.005}$ $\times$ 0.144 $\pm{0.005}$ & 105 $\pm{20}$ & 51.68 $\pm{0.39}$ & 82.73 $\pm{0.94}$ & 5.03 \\
19 & DoAr 33 & II &16:27:39.004 -23:58:19.149& 0.225 $\pm{0.006}$ $\times$ 0.176 $\pm{0.005}$ & 78.2 $\pm{5.6}$ & 37.68 $\pm{0.49}$ & 76.4 $\pm{1.4}$ & 4.64 \\
20 & GY 314 & II &16:27:39.422 -24:39:15.940& 0.258 $\pm{0.005}$ $\times$ 0.145 $\pm{0.006}$ & 138.9 $\pm{2.1}$ & 72.12 $\pm{0.86}$ & 151.9 $\pm{2.5}$  & 9.23  \\
21$^{a}$a & SR 9A & II &16:27:40.275 -24:22:04.568& 0.073 $\pm{0.016}$ $\times$ 0.065 $\pm{0.021}$ & 69 $\pm{83}$ & 13.15 $\pm{0.26}$ & 14.83 $\pm{0.48}$ & 0.90  \\
21$^{a}$b &  SR 9B & II &16:27:40.272 -24:22:03.888& -- $\times$ -- & ... & 3.02 $\pm{0.036}$ &3.02 $\pm{0.036}$ & 0.18\\
22 & SR 20 W & II &16:28:23.337 -24:22:41.070& 0.420 $\pm{0.022}$ $\times$ 0.145 $\pm{0.011}$ & 65.7 $\pm{1.7}$ & 22.16 $\pm{0.76}$ & 64.1 $\pm{2.9}$ & 3.90 \\
23$^{b,*}$Aab & EM* SR 13Aab & II &16:28:45.266 -24:28:19.358& 0.412 $\pm{0.040}$ $\times$ 0.329 $\pm{0.034}$ & 90 $\pm{27}$ & 31.7 $\pm{2.2}$ & 148 $\pm{12}$ & 9.00 \\
23$^{b}$ & EM* SR 13B & II &16:28:45.28 -24:28:19.318& ...  & ... & ... &$<$ 0.85 &$<$ 0.04 \\
24 & WSB 63 & II & 16:28:54.071 -24:47:44.694 & 0.266 $\pm{0.005}$ $\times$ 0.107 $\pm{0.006}$ & 0.07 $\pm{1.28}$ & 15.53 $\pm{0.17}$ & 30.55 $\pm{0.48}$ & 1.86 \\
25 & WSB 67 & II &16:30:23.398 -24:54:16.511& 0.175 $\pm{0.011}$ $\times$ 0.112 $\pm{0.013}$ & 12.8 $\pm{8.4}$ & 10.83 $\pm{0.21}$ & 16.82 $\pm{0.49}$ & 1.03 \\
26$^{a}$a & ROXS 42Ca & II &16:31:15.738 -24:34:02.487&  -- $\times$ --& 115.6 $\pm{3.7}$ & 3.96 $\pm{0.16}$ & 4.14 $\pm{0.18}$ & 0.25\\
26$^{a}$b & ROXS 42Cb & II &16:31:15.748 -24:34:02.72&  -- $\times$ --&  ...  &  $<$ 0.89 & $<$ 0.89 & $<$ 0.4 \\
27$^{a}$a & DoAr 43a & II &16:31:30.873 -24:24:40.288& 0.267 $\pm{0.008}$ $\times$ 0.110 $\pm{0.006}$ & 38.3 $\pm{1.6}$ & 18.8 $\pm{0.28}$ & 36.29 $\pm{0.78}$ & 2.21 \\
27$^{a}$b & DoAr 43b & -&16:31:31.025 -24:24:37.484& 0.121 $\pm{0.021}$ $\times$ 0.111 $\pm{0.025}$ & 97 $\pm{85}$ & 4.83 $\pm{0.15}$ & 6.58 $\pm{0.32}$ & 0.40 \\
28 & 2MASS J16313124-2426281 & II &16:31:31.245 -24:26:28.438& 1.301 $\pm{0.029}$ $\times$ 0.157 $\pm{0.005}$ & 49.05 $\pm{0.21}$ & 14.75 $\pm{0.25}$ & 124.8 $\pm{2.4}$ & 7.60 \\
29$^{c}$  & DoAr44 & II &16:31:33.455  -24:27:37.515& 0.911 $\pm{0.147}$ $\times$ 0.821 $\pm{0.134}$ & 63 $\pm{59}$ & 12.4 $\pm{1.8}$ & 262 $\pm{41}$ & 15.99 \\
30 & 2MASS J16314457-2402129 & II &16:31:44.577 -24:02:13.475& 0.110 $\pm{0.007}$ $\times$ 0.069 $\pm{0.014}$ & 133.8 $\pm{8.9}$ & 15.30 $\pm{0.16}$ & 18.81 $\pm{0.32}$& 1.14 \\
31$^{b}$a & LDN 1689 IRS 5A & F &16:31:52.111 -24:56:16.030& 0.117 $\pm{0.013}$ $\times$ 0.113 $\pm{0.013}$ & 79 $\pm{84}$ & 69.9 $\pm{1.3}$ & 94.4 $\pm{2.7}$ & 5.75\\
31$^{b}$ba & LDN 1689 IRS 5Ba & -&16:31:51.929 -24:56:17.44& ... & ... & $<$ 1 & $<$ 1 & $<$ 0.06\\
31$^{b}$bb & LDN 1689 IRS 5Bb & -&16:31:51.915 -24:56:17.376& 0.129 $\pm{0.025}$ $\times$ 0.047 $\pm{0.037}$ & 117 $\pm{17}$ & 4.39 $\pm{0.16}$ & 5.47 $\pm{0.33}$ & 0.34 \\
32 & WSB 74 &- &16:31:54.700 -25:03:24.000&-- $\times$ -- & ... & ... & $<$ 0.71 & $<$ 0.04 \\
33$^{a}a$ & DoAr 51A &- &16:32:11.848 -24:40:21.90&-- $\times$ -- & ... & ... & $<$ 0.75 & $<$ 0.05 \\
33$^{a}b$ & DoAr 51B &- &16:32:11.904 -24:40:21.76&-- $\times$ -- & ... & ... & $<$ 0.75 & $<$ 0.05 \\
34$^{a}$a& L1689-IRS 7A & II &16:32:21.047 -24:30:36.309& 0.080 $\pm{0.012}$ $\times$ 0.066 $\pm{0.021}$ & 144 $\pm{66}$ & 29.17 $\pm{0.5}$ & 33.43 $\pm{0.96}$ & 2.03 \\
34$^{a}$b &  L1689-IRS 7B  & II &16:32:20.811 -24:30:29.487& 0.110 $\pm{0.015}$ $\times$ 0.066 $\pm{0.038}$ & 156 $\pm{28}$ & 5.16 $\pm{0.13}$ & 6.32 $\pm{0.25}$ & 0.39 \\
35 & Haro 1-17 & II &16:32:21.928 -24:42:15.208& 0.135 $\pm{0.016}$ $\times$ 0.064 $\pm{0.023}$ & 79 $\pm{12}$ & 6.97 $\pm{0.2}$ & 8.87 $\pm{0.41}$ & 0.53 \\
36$^{a,*}$  & 2MASS J16335560-2442049AB & II &16:33:55.610 -24:42:05.370& 0.670 $\pm{0.084}$ $\times$ 0.461 $\pm{0.059}$ & 77 $\pm{15}$ & 24.9 $\pm{2.7}$ & 233 $\pm{28}$ & 14.17 \\
38$^{c,d}$  & WSB 82 & II &16:39:45.440 -24:02:04.250& 1.301 $\pm{0.057}$ $\times$ 0.632 $\pm{0.028}$ & 171.6 $\pm{2.2}$ & 17.44 $\pm{0.72}$ & 437 $\pm{19}$ & 26.52 \\
39 & 2MASS J16214513-2342316 & I &16:21:45.122 -23:42:32.182& 0.628 $\pm{0.032}$ $\times$ 0.118 $\pm{0.014}$ & 174.33 $\pm{0.97}$ & 21.73 $\pm{0.81}$ & 94.3 $\pm{4.3}$ & 5.73\\
40 & 2MASS J16263682-2415518a  & F &16:26:36.827 -24:15:52.298& 0.390 $\pm{0.046}$ $\times$ 0.323 $\pm{0.041}$ & 6.1 $\pm{28.3}$ & 10.17 $\pm{0.83}$ & 46.3 $\pm{4.5}$ & 2.82\\
41 & WL6  & I &16:27:21.791 -24:29:53.826& 0.106 $\pm{0.019}$ $\times$ 0.072 $\pm{0.027}$ & 16 $\pm{28}$ & 15.33 $\pm{0.37}$ & 18.73 $\pm{0.73}$ & 1.14 \\
42 & GY 312 & I &16:27:38.936 -24:40:21.058& 0.390 $\pm{0.009}$ $\times$ 0.121 $\pm{0.007}$ & 168.27 $\pm{0.94}$ & 34.45 $\pm{0.56}$ & 92.9 $\pm{2}$ & 5.66 \\
43 & 2MASS J16274161-2446447 & I &16:27:41.601 -24:46:45.082& 0.267 $\pm{0.011}$ $\times$ 0.139 $\pm{0.009}$ & 99.8 $\pm{3.1}$ & 14.75 $\pm{0.32}$ & 31.3 $\pm{0.95}$ & 1.90 \\
44 & GY 344 & - &16:27:45.800 -24:44:54.000& -- $\times$ -- & ... & ... & $<$ 0.74 & $<$ 0.04\\
45$^{a}a$ & YLW 52a & F &16:27:51.796 -24:31:46.048& 0.216 $\pm{0.043}$ $\times$ 0.106 $\pm{0.066}$ & 129 $\pm{17}$ & 3.13 $\pm{0.28}$ & 5.46 $\pm{0.72}$ & 0.34 \\
45$^{a}b$ & YLW 52b & F &16:27:51.479 -24:31:40.33& ... & ...  & $<$ 0.7  &  $<$ 0.7  & $<$ 0.03   \\
46 & WSB 60 & II &16:28:16.503 -24:36:58.463& 0.554 $\pm{0.026}$ $\times$ 0.512 $\pm{0.025}$ & 135 $\pm{27}$ & 26.31 $\pm{0.98}$ & 232.2 $\pm{9.6}$ & 14.17 \\
47 & 2MASS J16313383-2404466 & F &16:31:33.831 -24:04:47.036&$<$ 0.16 $\times$ 0.057 & ...& 4.4 $\pm{0.24}$ & 5.52 $\pm{0.48}$ & 0.34 \\
48 & IRS 63 & F &16:31:35.659 -24:01:29.893& 0.521 $\pm{0.024}$ $\times$ 0.359 $\pm{0.018}$ & 150 $\pm{5.2}$ & 123.5$\pm{4.8}$ & 776 $\pm{35}$ & 47.19 \\
49 &2MASS J16442430-2401250 & - &16:44:24.300 -24:01:25.000& -- $\times$ --  & ... & ... & $<$ 0.80 & $<$ 0.05\\
50 & Haro 1-11 & II &16:27:38.325 -23:57:32.936& 0.151 $\pm{0.020}$ $\times$ 0.113 $\pm{0.021}$ & 84 $\pm{30}$ & 7.61 $\pm{0.28}$ & 11.21 $\pm{0.62}$ & 0.68 \\
\enddata
\begin{tablenotes}
      \small
      \item $^{a}${Field is a binary source}
       \item $^{b}${Field is a triple source}
        \item $^{c}${Transition Disk}
        \item $^{d}${Evidence of gap in disk}
        \item ${*}${Circumbinary Disk}
         \item ${*,?}${Potential circumbinary Disk}
       \end{tablenotes}  
\label{tab:source}
\end{deluxetable}
\end{longrotatetable}
\section{Discussion}  \label{sec:dis}
\subsection{Comparison with the Taurus-Auriga Molecular Cloud}
In this work, we have used ALMA to map the distribution of 870 $\mu$m emission from 49 selected
pre-main sequence stellar systems in the 
$\rho$ Ophiuchus molecular cloud and used these maps to construct the distribution
 of disk fluxes and radii from various subpopulations. A natural question to ask is how the systems in 
 one molecular cloud compare to those of another. To do this, we have compiled a 
 target list of sources in the Taurus-Auriga molecular cloud to which we compare our sample. 
 Taurus represents an obvious choice for such a comparison. First, it has a well-characterized stellar 
 population and disk population due to its proximity (145 pc; \citealt{lo07,to07,to09})
 as well as a relatively uniformly low extinction across the whole cloud (\citealt{lo10}). Second, $\rho$ Oph has 
 a relatively low stellar density across much of its volume and very few UV/X-ray luminous O/B-type stars,
 much like Taurus and opposed to clusters such as 
 Orion. Such environmental impacts are known to have severe and deleterious effects on protoplanetary disk masses and radii 
 (e.g., \citealt{ma14}). Finally, the two clusters are close to the same age: $\rho$ Oph is between 0.5 - 2 Myr \citep{wi08} old, 
 while Taurus is in the vicinity of 1-2 Myr old (\citealt{lu10}). 
 
In order to quantify how common our sample is, we have constructed a sample of 
Taurus sources to which we compare our Oph sample. To do this, we used the results of the 
\textit{Spitzer} survey performed by \citet{re10}. They surveyed approximately 44 square
degrees of Taurus in each of the 7 different IRAC/MIPS bands. To ensure that our comparison stars
were in Taurus, we restricted our selection to the subsample of their survey that had already 
previously been identified 
as Taurus members, rather than those sources that were inferred to be Taurus members based on colors from their 
survey. As in our survey, we only included sources with detections in all of the IRAC bands as well as the 24 and 
70 $\mu$m bands in the same fashion as was done for our present survey. The sensitivities of the \citet{re10} survey are similar to those of the c2d survey, so this is 
probably a fair comparison. After selecting candidate sources in Taurus, we restricted the sample to those sources 
that had (sub)-millimeter flux information in the literature. 
Where multi-band photometry was available, we used the derived spectral index to infer the  870 $\mu$m flux 
density; where it was not available, we assumed that the intrinsic spectral index was 3. The qualitative results for this work do not depend on the precise value of $\alpha$ we assume. We use the same KM estimators to compare the corresponding subpopulations of Taurus-Auriga objects with Oph objects. 
The $p$-values for these comparisons are found in Table 4.

\begin{figure*}[htp]
\centering
\begin{tabular}{cc}
\includegraphics[scale=0.6]{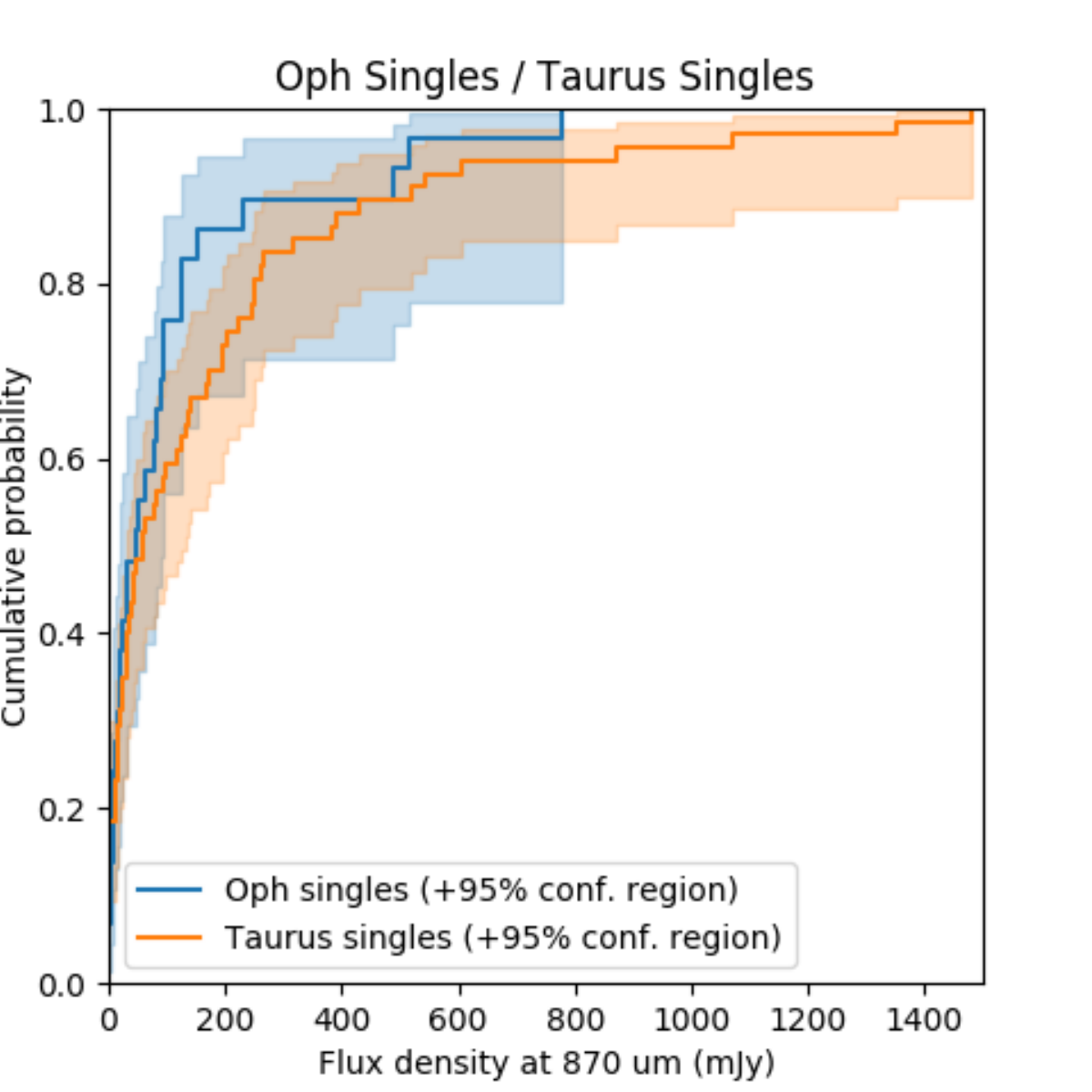}
\includegraphics[scale=0.6]{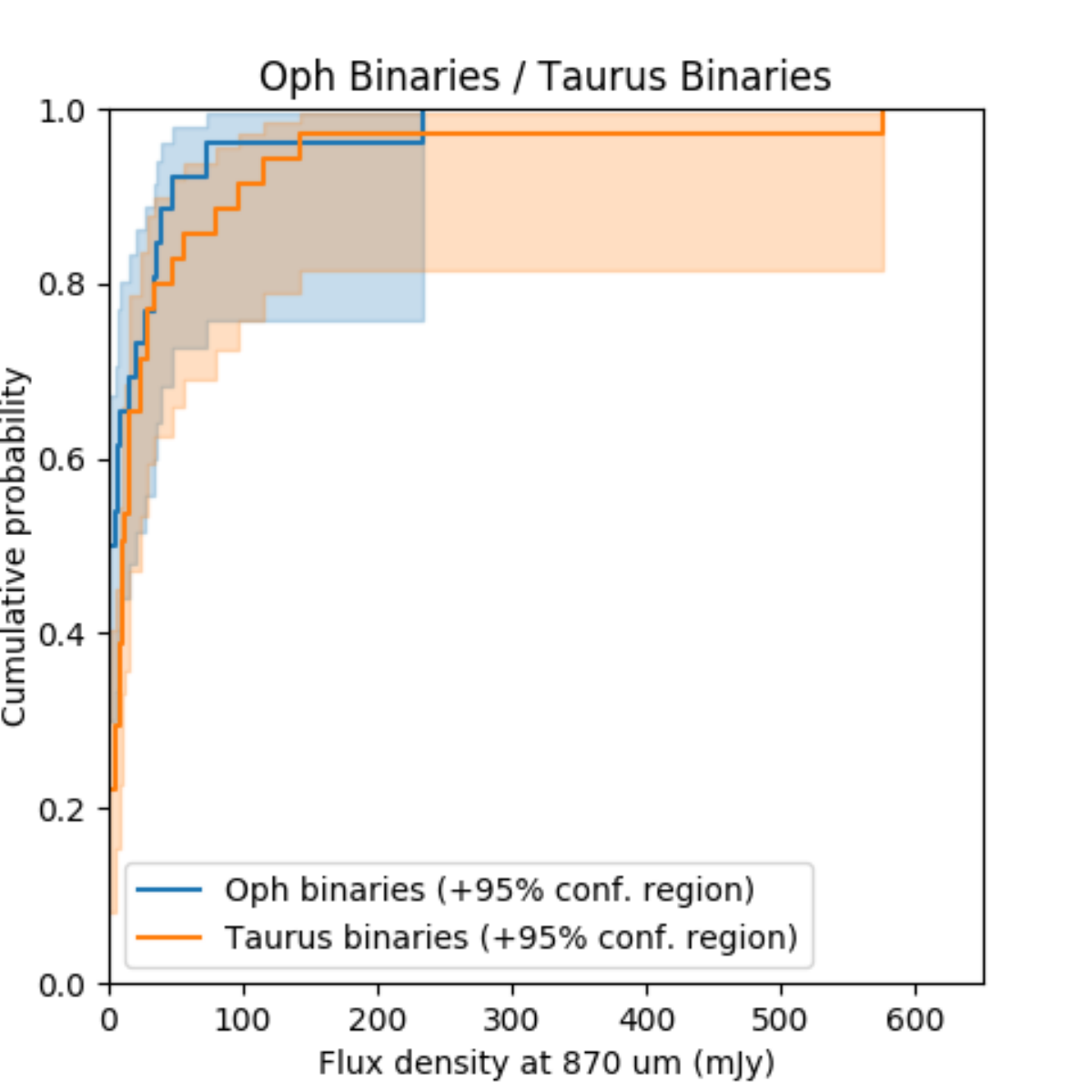}\\
\includegraphics[scale=0.6]{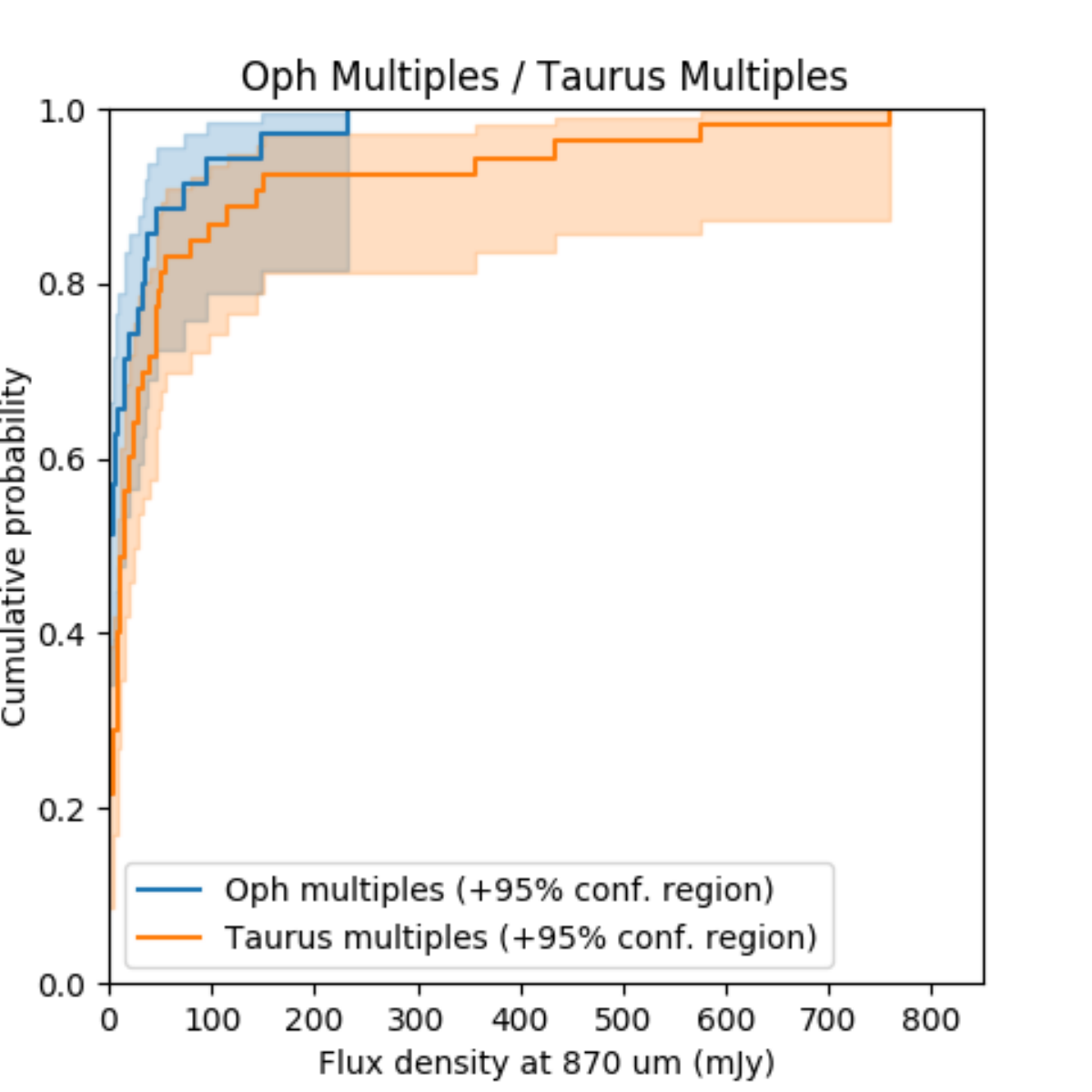}
\includegraphics[scale=0.6]{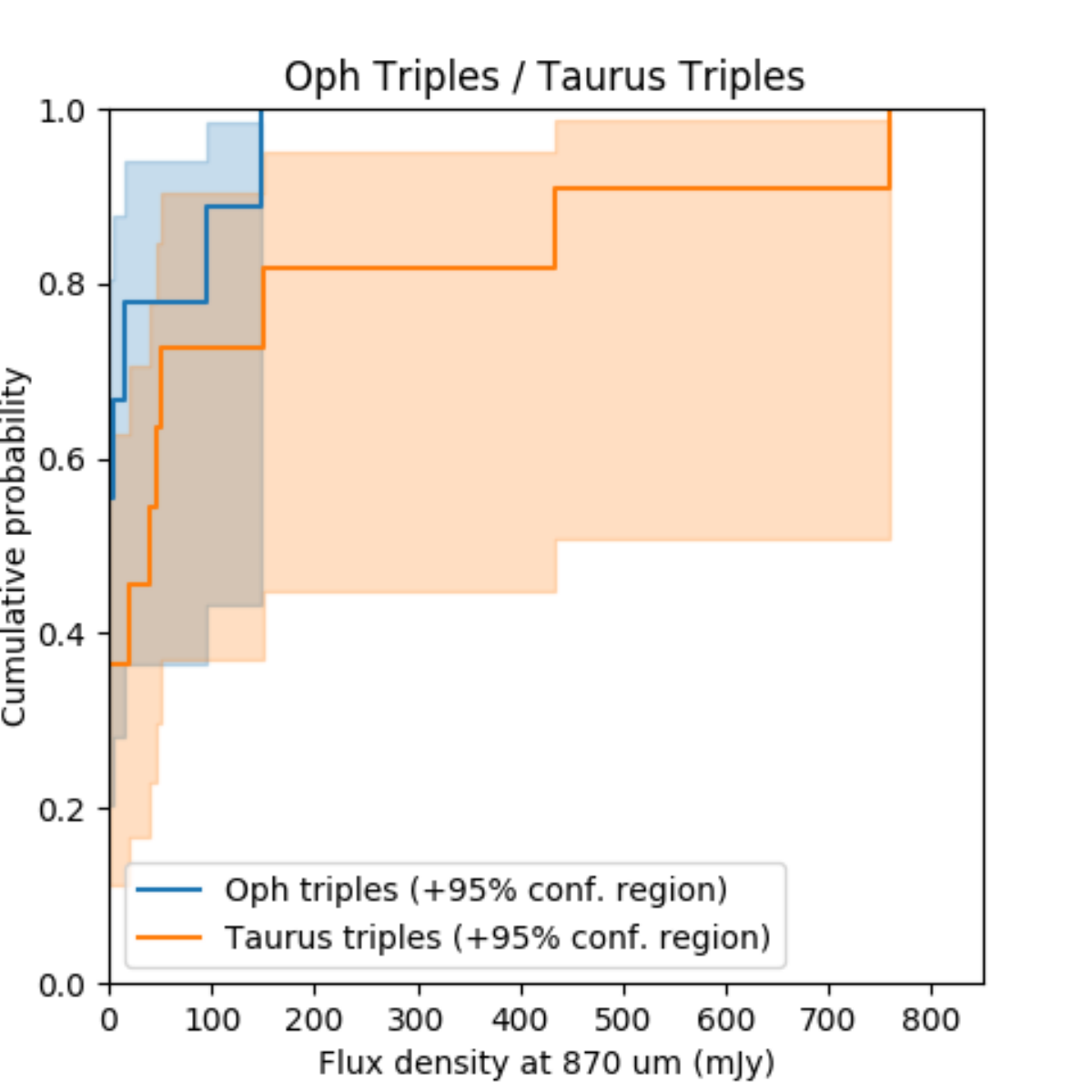}
\end{tabular}
\caption{CDF comparisons of flux in $\rho$ Ophiuchus and Taurus for isolated protostars, binaries, triples and multiples (binaries plus triples).}
\label{fig:taupop}
\end{figure*}

\begin{figure*}[htp]
\centering
\begin{tabular}{cc}
\includegraphics[scale=0.6]{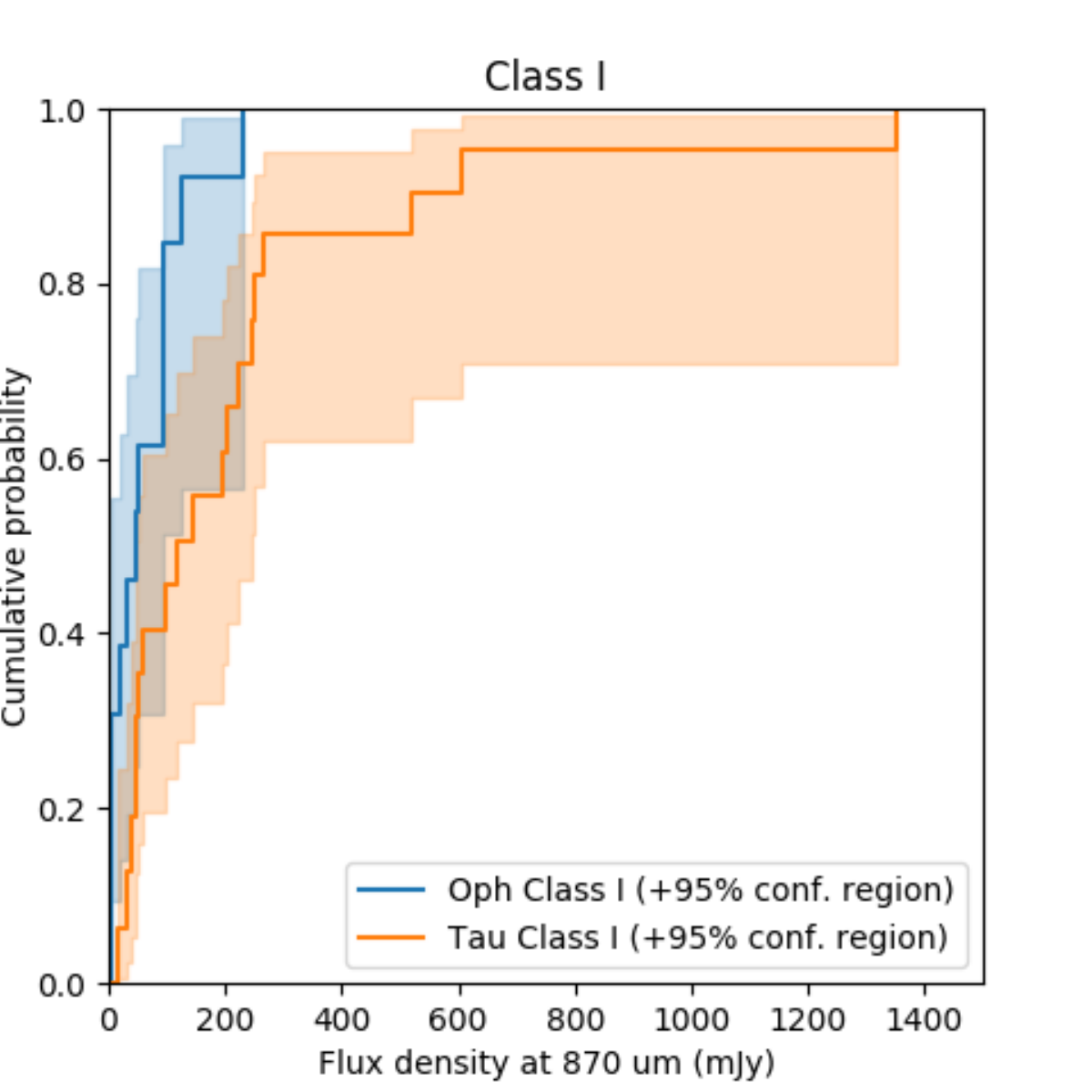}
\includegraphics[scale=0.6]{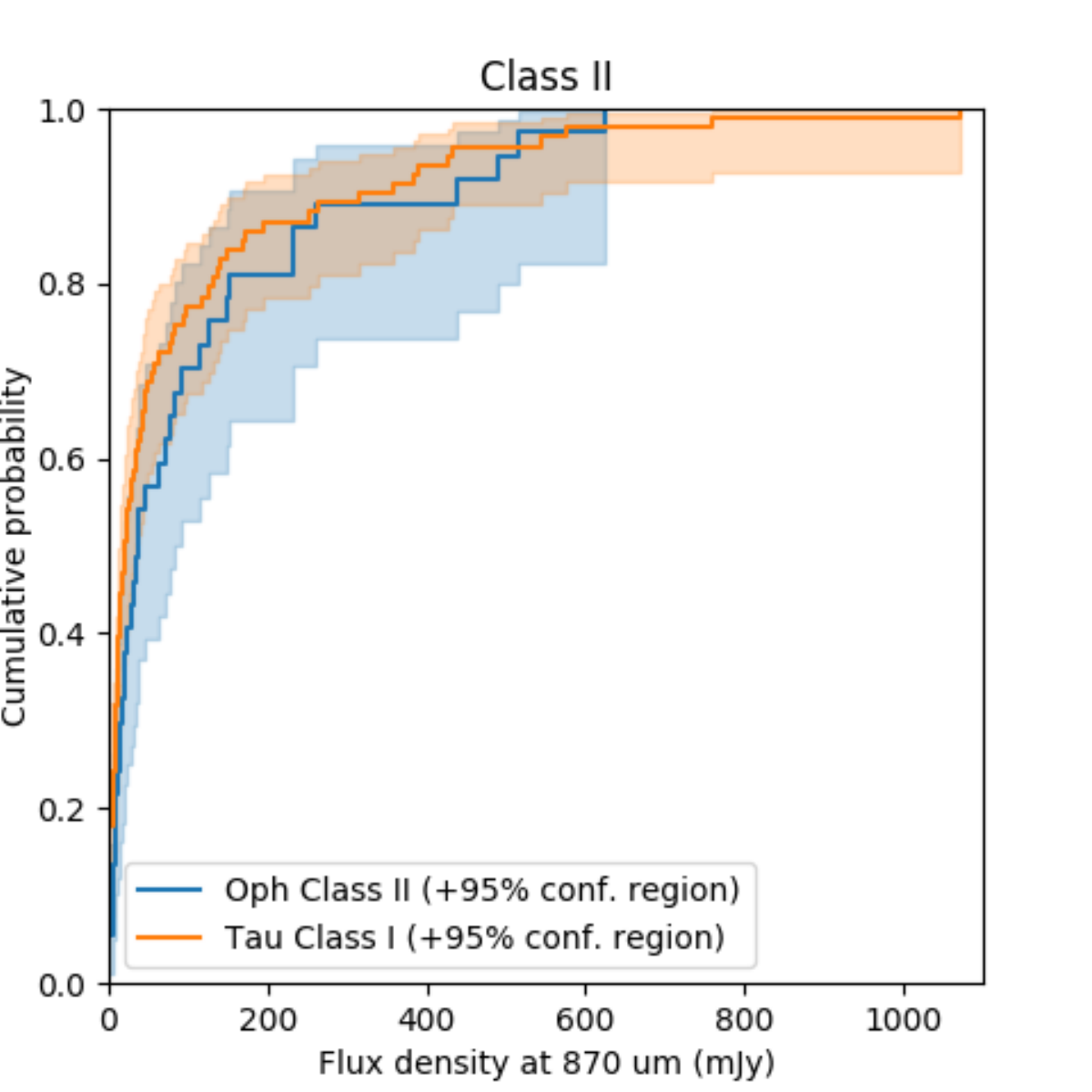}
\end{tabular}
\caption{CDF comparisons of flux in $\rho$ Ophiuchus and Taurus for Class I and Class II protostars.}
\label{fig:tauclass}
\end{figure*}

In Figures \ref{fig:taupop} and \ref{fig:tauclass} we show the CDF comparisons of the Oph and Taurus 
populations. We find that the single sources have different median fluxes (31 mJy vs. 57 mJy), with their 
corresponding low $p$-value (0.00282), most likely due to the Taurus population having a high flux tail in its distribution. 
One possibility for this dichotomy is the difference in the environments between the two clouds. $\rho$ Oph tends to have more clustered YSOs while Taurus's YSOs are more dispersed.
We find that Ophiuchus typically has dimmer binary and triple systems, as well as Class II protostars, with
its Class I population being much dimmer ($\sim$31 mJy vs. $\sim$116 mJy) than that of Taurus. Due to the low
number of Class I YSOs in our survey, this is likely due to small number statistics. 

\begin{table}[h!]
\begin{center}
\caption{Statistics of various comparisons.}
\vspace{0.2cm}
\begin{tabular}{|r|r|}
\hline
Comparison & Flux $p$-value \\ \hline
Singles & 0.00434\\
Binaries & 0.70718\\
Triples &  0.18367\\
Multiples & 0.23964\\
Class I & 0.00122\\
Class II & 0.50860\\ \hline \hline
Taurus Population & Median Flux (mJy) \\ \hline
Singles & 57.4\\
Binaries & 10.7\\
Triples & 29.2\\
Multiples & 12.9 \\
Class I Sources & 115.8\\
Class II Sources &  21.4\\ \hline
\end{tabular}  
\end{center}
\scriptsize Comparison of the various $p$-values obtained from each CDF. Note that 'multiples' is a combination of both binary and triple systems.
\label{tab:taur}
\end{table}

\subsection{Disks in Binary Systems and Tidal Truncation}
Protoplanetary disks in binary systems are subject to far more interactions than disks in single systems, due
to the manner in which disks around stars and stellar companions interact. The disks surrounding these protostars can only grow to a certain 
radius before that material is stripped away by its companion. This is likely
due to the interactions with their companions, yielding a loss of disk material \citep{je96,ha12}. The lower disk fluxes can be interpreted as being
due to lower disk masses. Theory indicates that disk truncation in binaries is particularly sensitive to the binary's 
semimajor axis $a$ and eccentricity $e$. Essentially, the closer the periastron distance $d = a(1-e)$, the more severe the truncation. 
We use the analytic model described in \citet{pi05} to estimate
\begin{figure*}[htp]
\centering
\begin{tabular}{cc}
\includegraphics[scale=0.6]{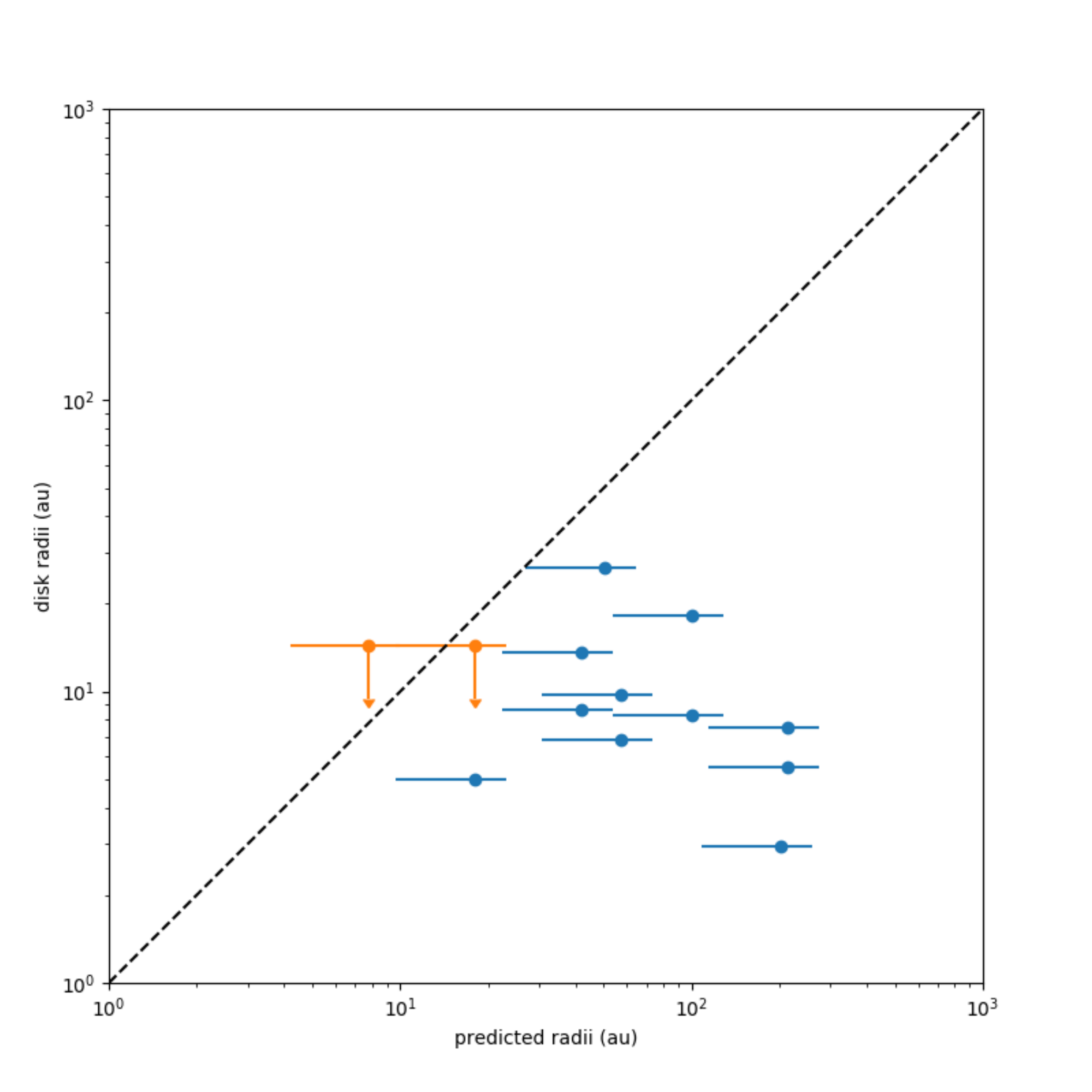}
\end{tabular}
\caption{Measured disk radii in the binary systems observed (see Table \ref{tab:source}) compared with the expected disk radii based on tidal interaction models from \citet{pi05}.
Note that the expected disk radii are lower than the equality line, meaning that disk truncation is not setting the disk radii in $\rho$ Ophiuchus. The orange arrows represent the two sources for which we have upper limits on the radius. The error bars on the points represent the 68\% confidence region.}
\label{fig:trunc}
\end{figure*}
the equilibrium truncated radius of our binary sources. This model yields a prediction for a circumstellar disk's truncation radius given its host binary's orbital elements
 $a$ and $e$, as well as the mass ratio $q$, which we assume to be unity (this has little effect, as the truncation radius depends only 
 very weakly on $q$ for reasonable values of $q$).  Because we have no orbital information on our binary systems outside of a projected 
 separation on the sky, we implement a statistical method to estimate the true orbital elements $a$ and $e$ based on the projected separation of the two stars; for details, see \citet{ha12}. We then convert this to a prediction for the tidal radii. The detailed predictions are somewhat 
sensitive to the choice of the eccentricity probability density function; we choose a uniform distribution picking $e$ between 0 and 1 for this.  
As seen in Figure \ref{fig:trunc}, all of our sources, barring the two with
upper limit detections, are well below the equilibrium line. This means that for the binary systems we observed in Oph, truncation
is not responsible for the disk size observed. This is in contrast to what \citet{ha12} (see their Figure 11) found for the 
Taurus binary systems. The Taurus systems have a much more scattered distribution
with points both above and below the equilibrium.

Figure \ref{fig:trunc} shows that the measured dust disk radius and that 
predicted from our statistical modeling disagree. However, there are two caveats
to this analysis. First, the gas and dust extents are not necessarily the same. 
Dust-size dependent aerodynamic effects such as radial drift can lead to 
differences in the structure of the gas (which comprises the bulk of the disk 
mass) and that of the large particles responsible for the millimeter continuum 
emission  (e.g., \citealt{we77,pe12}). Due to these effects, the dust 
emission extent is expected to be more compact than the gas-line emission, with
theoretical estimates of the ratio of 0.88 mm continuum extent to CO emission 
line extent ranging between 1.5 to about 4 (e.g., \citealt{fa17}). Observational
evidence also suggests this to be the correct range (e.g., \citealt{an12,va17}).
Accordingly, the measured radii could be corrected by a typical correction 
factor of $\sim 2-3$ and be brought into good agreement with truncation models. 
Alternatively, because our predictions for the tidal radii are dependent on the 
(unknown) eccentricity distribution for pre-main sequence binaries, it 
is plausible that an eccentricity distribution weighted more towards moderate
 to high eccentricity would alleviate the discrepancy we note. For main-sequence
 stars with periods $P \gtrsim$ 100 days, observations are consistent with a 
uniform distribution between 0 and 1 \citep{du13}.  It is plausible that, in the 
past, the progenitors of these systems (and the analog of the disk-bearing 
systems we focus on here) had higher eccentricities that were subsequently
damped due to star/disk interaction (e.g., \citealt{al01}), making the 
higher eccentricity distribution the more appropriate one to use here.

\subsection{Transition and Gapped Disks}
An interesting outcome of this ALMA survey is how diverse the YSO 
population we observed is. As discussed in \S \ref{sec:sam}, our aim was to 
probe more evolved protostars, to characterize their disks. Of the 49 stellar systems we observed, 5 include transition disks (see Figure \ref{fig:trans}). Our ALMA observations were only $\sim$36 seconds and provide 
unprecedented detail in all 5 of these sources. Three of these are known transition disks that have
been heavily observed and studied in both the infrared and the sub-/millimeter regimes (ROph 13, ROph 29, ROph 36). One source, ROph 3, does not have existing sub-/millimeter observations,
but was determined to be a transition disk from IR data. Finally, ROph 38 has no existing millimeter data, and, unlike the other transition disks observed in this survey, there is no indication of a central cavity in the broadband \textit{Spitzer} near to mid-infrared photometry taken during the C2D survey \citep{ev03}. We, however, detect a large millimeter cavity as well as a gap and a ring-like structure of low-level emission surrounding it. This indicates that, while the central cavity maybe devoid of mm-sized particles, it is \textit{not} devoid of small particles.

The detection of disks that show evidence for narrow gaps in their emission is a particularly exciting result from our survey. Such gaps in the 
millimeter emission from the disk have been directly 
imaged previously in the young Class I/II object HL Tau \citep{al15} and the nearby older Class II TW Hya \citep{no16,an16}, as well as in the higher mass
Herbig Ae stars HD 163296 \citep{is16} and HD 169142 \citep{fe17}. Modeling 
of ALMA continuum data at 0.87 and 1.3 mm of the young 
Class II star AA Tau also suggests multiple gaps in this star's disk 
\citep{lo17}. The leading candidates for how the gaps open are either a forming protoplanet/gas-giant core gravitationally 
torques material around it, effectively repelling some disk material away from it \citep{lp86}, or through enhanced grain growth due to pressure bumps caused by
planets \citep{bi10}.
Other suggestions from theorists for forming rings and gaps the millimeter emission include dust sintering \citep{oz16} and disk surface density variation driven by 
inhomogeneous magnetic field distribution (e.g., \citealt{fl15}) or magnetic disk-winds \citep{su17}.
The exact details of the gap opening, including the gap structure's 
dependence on planetary embryo mass and surrounding disk structure, 
have not been fully analytically described \citep{cr06}. 
In fact, it is uncertain whether a single planet per gap is required for 
gap formation or if a single planet can carve multiple gaps \citep{do17}.
It is, however, generally agreed that higher embryo masses carve more 
substantial gaps. Numerical calculations indicate that a range of planetary masses
$\gtrsim 0.2 M_{\mathrm{Jupiter}}$ can carve observable gaps in 
disks' millimeter emission. 

\begin{figure*}[htp]
\centering
\includegraphics[width=\linewidth]{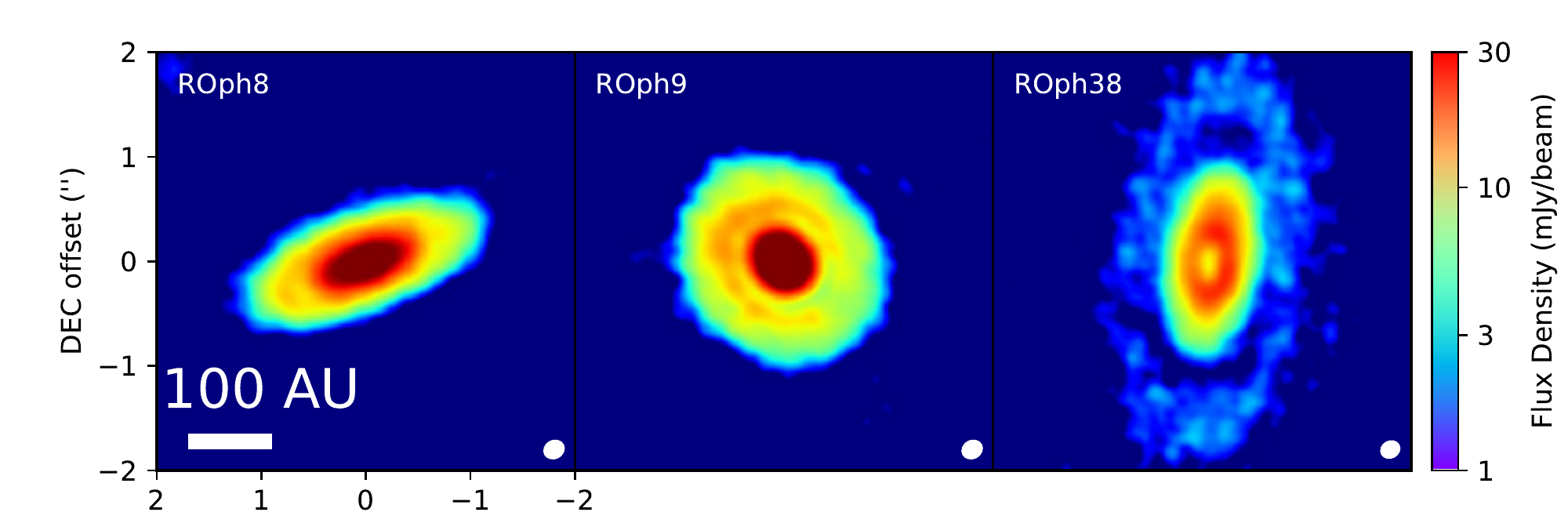} 
\caption{ROph 8, ROph 9, and ROph 38, the candidate gapped disks imaged in our survey. Note that the color-scale is saturated so as to more clearly show the gaps.}
\label{fig:gaps}
\end{figure*}

\begin{figure*}[htp]
\centering
\begin{subfigure}
\centering
\includegraphics[width=0.31\linewidth]{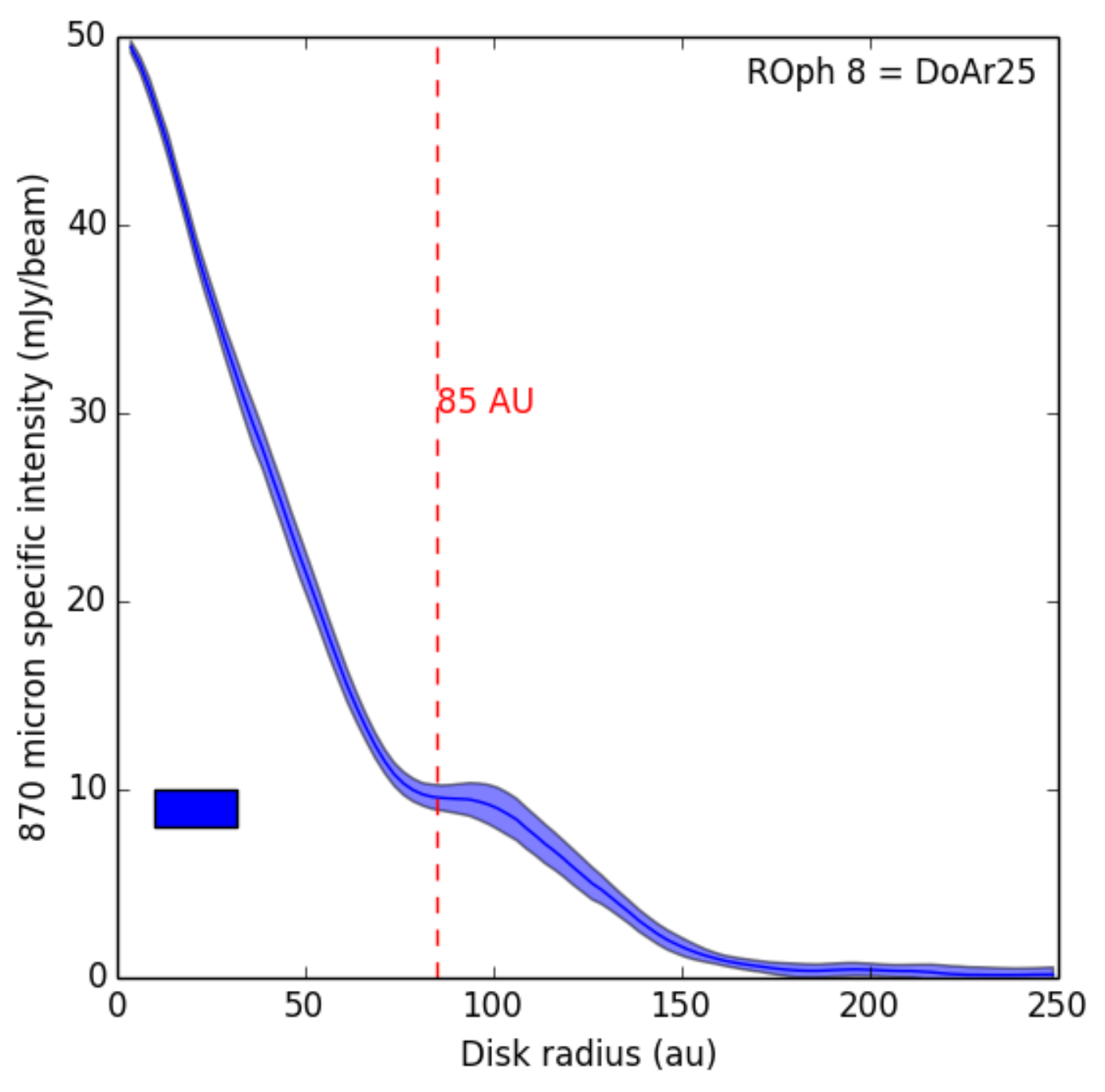}    
\end{subfigure}
\begin{subfigure}
\centering
\includegraphics[width=0.31\linewidth]{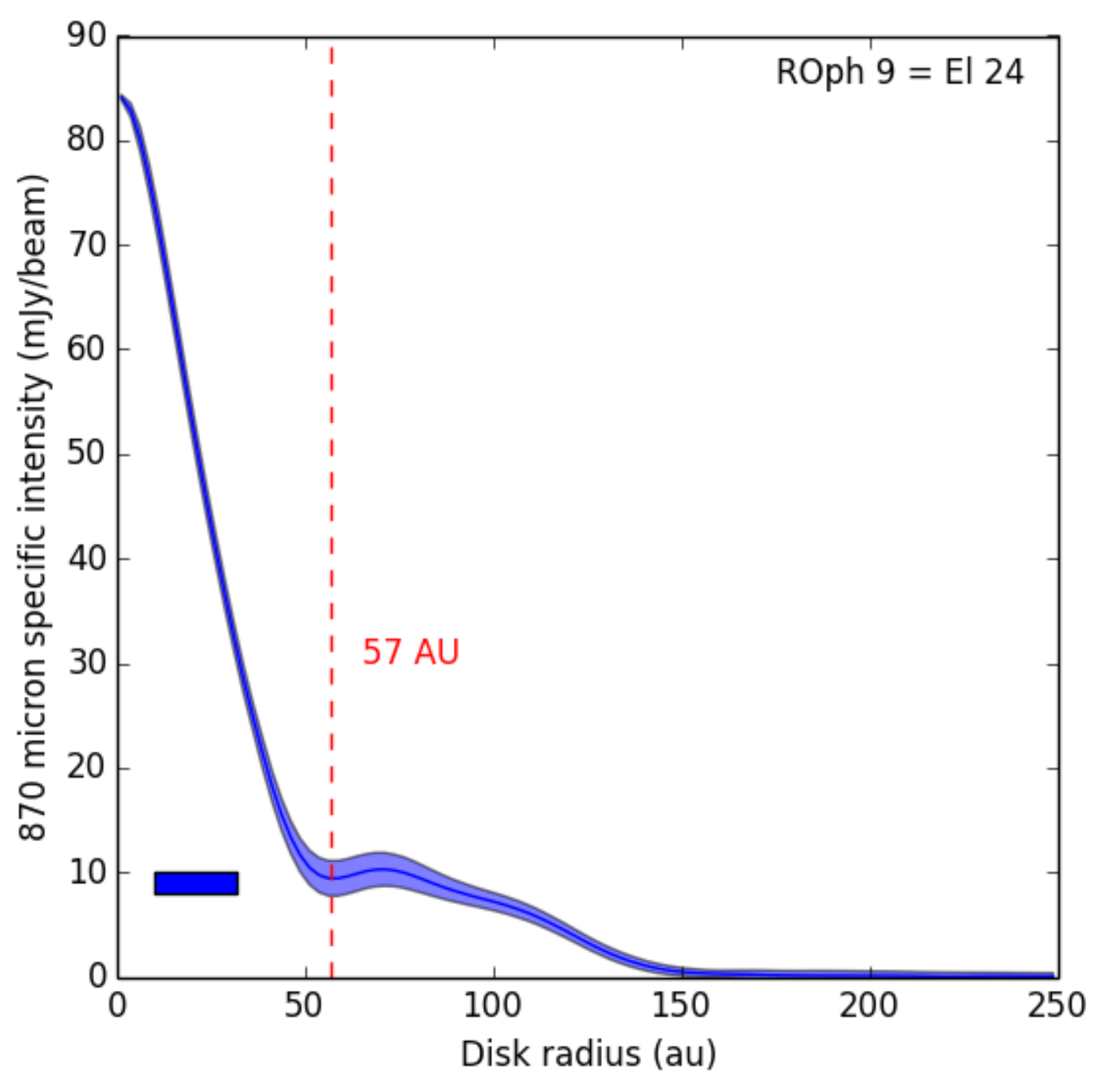}    
\end{subfigure}
\begin{subfigure}
\centering
\includegraphics[width=0.31\linewidth]{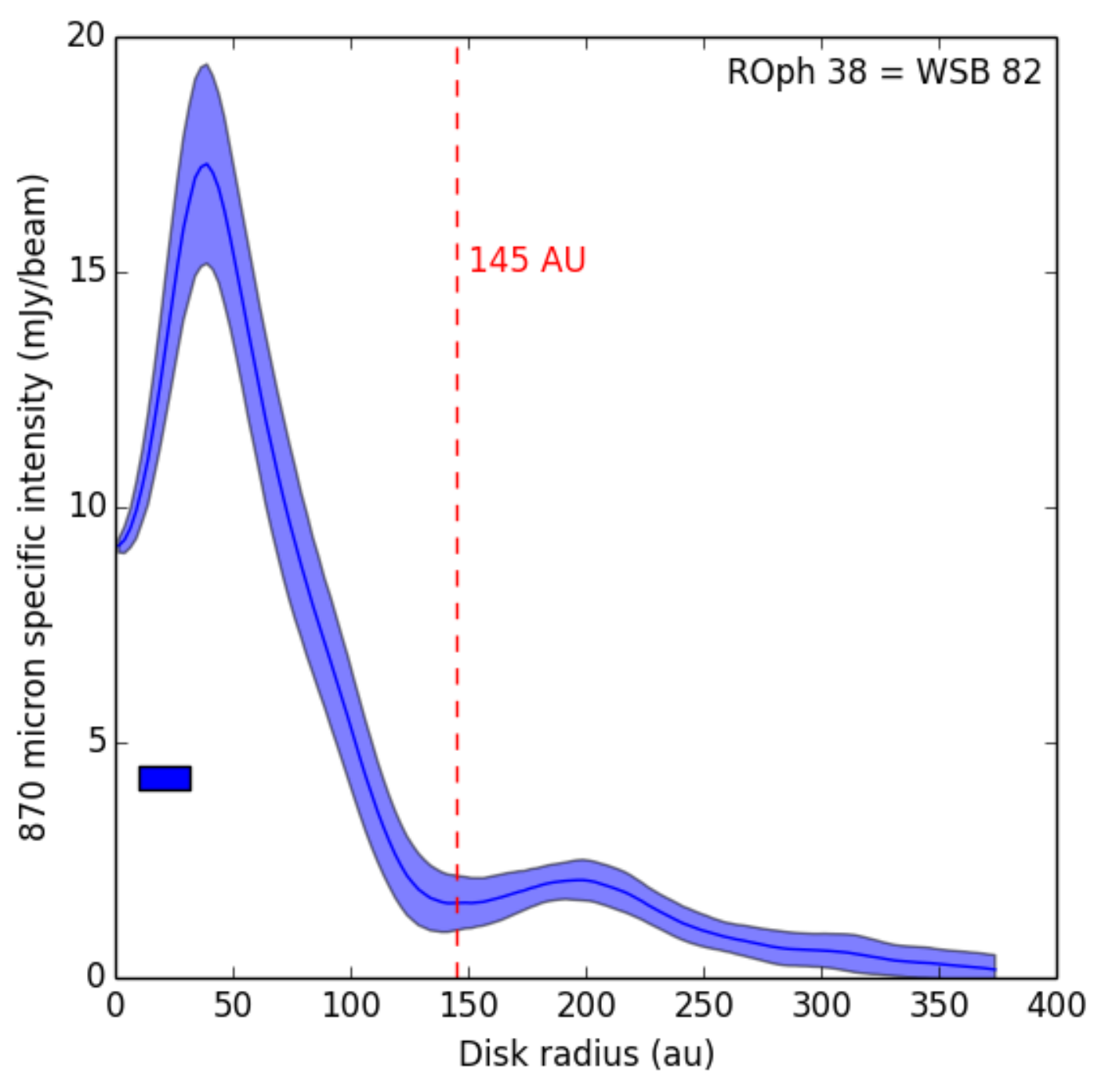}    
\end{subfigure}
\caption{Deprojected, azimuthally averaged surface brightness profiles for three sources in our survey with evidence for gaps. For ROph 9 and 38, the entire 
2$\pi$ in azimuth is averaged, whereas for ROph 8, only the region within 20 degrees of the disk major axis is averaged (due to the high disk inclination, 
the deficit in emission is seen only along the major axis). The resolution is shown as a thick horizontal bar. The estimated locations of the gaps are shown by red dashed vertical lines. }
\label{fig:profiles}
\end{figure*}

In our sample, we find two sources, ROph 9 (Elias 24) and ROph 38 (WSB 82) 
that exhibit clear evidence in the images of substantial disk gaps (see Figures \ref{fig:singles} and 
\ref{fig:trans}), while 
another source, ROph 8 (DoAr 25; see Fig \ref{fig:singles}) shows some evidence of a potential 
gap in the disk in its image. We present these sources again in Figure \ref{fig:gaps} with an altered color-scale to emphasize the gaps and low-lying emission in each disk. To quantify the structure of the gaps, we follow the procedure used by \cite{al15} to study the gap structure in the millimeter emission of HL Tau and deprojected each image 
using the fit disk center, inclination, and position angle, and produce azimuthally averaged 
surface brightness profiles. These profiles are shown in Figure \ref{fig:profiles}.

In the case of ROph 9 and 38, there are obvious deficits of emission observed at approximately 65 and 170 au, respectively. These gaps appear to be either unresolved or 
only marginally resolved by the synthesized beam of the array. There is a hint of a plateau
in the profile for ROph 8, which when combined with the imaging results, suggest a potential deficit in emission at approximately 95 au. Because the gaps are only marginally
resolved, they must be less than about 10 au in annular extent. To ensure that we are not `missing' sources that may have gaps that are not obvious in the images, we constructed 
these deprojected profiles for each source in our sample. Each source without obvious evidence for a millimeter cavity (see Fig. 4) shows a monotonically decreasing flux density with 
radius until the flux density starts to approach the noise level in the images.
We will present a more in depth analysis of both the transitional disks and the gapped disks in a future work. 
  
One potentially interesting question we can begin to ask is the fraction of disks ($f$) that show ongoing, present-day evidence for planet formation. 
If we consider either disk gaps or large millimeter cavities (in the 
absence of other explanations, such as known binarity) as evidence
of ongoing planet formation, we find that 6 out of 49 disks in our sample show evidence of forming planets that are massive enough to open up large gaps or cavities at the current epoch. This yields an 
estimate of $f = 0.122$ with a 95\% confidence interval of  $0.031 < f < 0.21$. Note that $f$ represents the fraction of systems that are estimated to have large ($\gtrsim 0.2 M_{\mathrm{Jupiter}}$) mass reservoirs that also have signposts of planet formation (i.e., gaps or central cavities).

\subsection{Asymmetric Dust Disks}

Asymmetries in the millimeter continuum emission from circumstellar disks have recently become of interest due to their likely origin in dust traps that may enable rapid grain growth past the barriers that, e.g., radial drift may impose \citep{pi11,ra17,mi17}. 
\begin{figure*}[htp]
\centering
\includegraphics[width=0.5\linewidth]{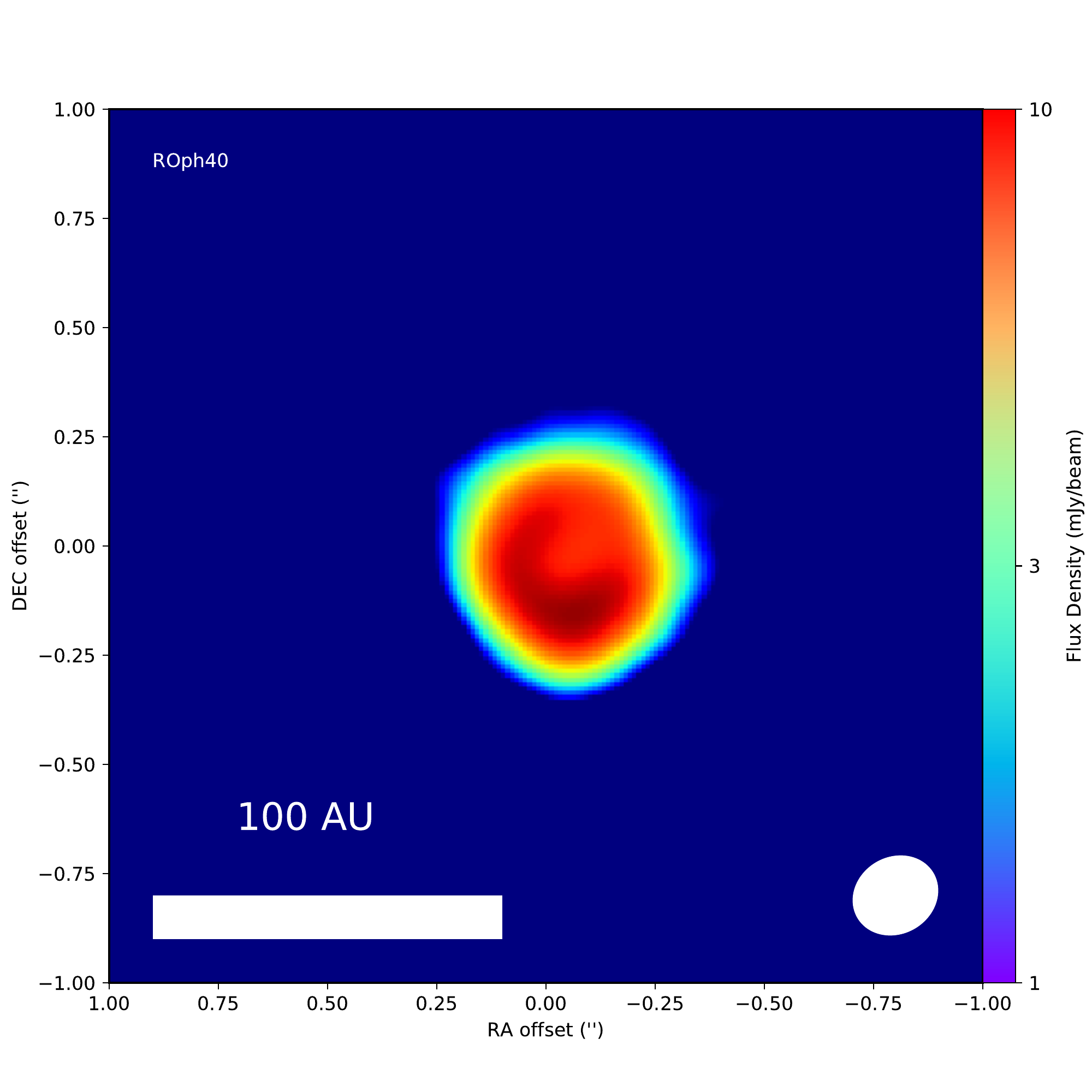} 
\caption{Map of the 870 micron continuum emission from the disk in the ROph 40 (ISO-Oph 51) binary system. The color-scale has been altered to emphasize the asymmetric dust emission and `horseshoe' shape in the emission.}
\label{fig:roph40}
\end{figure*}
These asymmetries are sometimes observed in transitional disks as a potential sign of a planet forming an azimuthally asymmetric pressure gradient within the surrounding disk (e.g., IRS 48, \citealt{vdm13}; SAO 206462 and SR 21, \citealt{pe14}). In each of our sample of transitional disks, we also observe somewhat substantial asymmetries in the outer disk, with a typical contrast of about 20\% from maximum to minimum in the profile of brightness vs azimuth as a given disk radius. In addition to the transitional disks we identify on the basis of a substantial millimeter cavity in
Figure \ref{fig:trans}, we also identify the primary disk in one of our binaries, ROph 40, as having a large asymmetry in its continuum emission. It is shown in Figure \ref{fig:roph40}. 
This source shows no infrared signature of being a transitional disk in either \textit{Spitzer} or \textit{Herschel} data. Furthermore, any cavity in the millimeter emission is not obvious, unlike in the analogous case of ROph 38. However, the asymmetry in the dust emission is reminiscent of what is observed in the transitional disks both we and others have mapped in the millimeter. Unfortunately, we do not have coverage of the requisite gas-line emission to be able to tell whether this is a true dust trap or a mere asymmetry in the overall mass distribution of the source. 
\section{Summary}  \label{sec:sum}
We have presented an ALMA imaging survey of the 870 $\mu$m dust continuum emission from the circumstellar
material of 49 systems in the $\rho$ Ophiuchus molecular cloud complex. These systems, having been 
selected on the basis of excess in each of the \textit{Spitzer} IRAC and MIPS bands, represent the stellar systems 
most likely to have sufficient circumstellar material to enable planet formation over the next few Myr. This survey, 
observing each source only for 36 seconds per source, shows the versatility and promise of the ALMA instrument 
for studies in star and planet formation. Many of these sources represent low-mass targets that have not been observed at millimeter-wavelengths before. 

We summarize our results and analysis below:

\begin{itemize}
	\item We divided the sources into several different populations (i.e., single stars, binaries, triple systems and transition disks) 
	and computed Kaplan-Meier product limit estimators to estimate the cumulative probability distribution for both disk 
	flux and disk radius for each population. We find significant differences in both flux and radius amongst the singles and 
	binaries in Oph: disk fluxes and radii in binaries are significantly smaller than in single stars. Similar results about the 
	fluxes have been noted previously (e.g \citealt{je94,je96,ha12}), but disk radii at millimeter wavelengths have never explicitly 
	been found to be smaller in disks in binaries compared to disks around isolated stars.  Large differences in circumstellar mass 
	(for which, assuming a single temperature and $\kappa$, flux can be a proxy for) and radius over a small range of ages illustrate 
	the diversity of conditions in the disk, wherein planets are forming.
	\item The lack of flux in the binary population is typically considered to be due to either disk truncation after formation, or caused 
	by something that sets the disk radii during formation. Using a statistical model to convert from projected separation to semi-major 
	axis and eccentricity, we computed the distribution of expected truncation radii using the analytic prescription of \cite{pi05} for 
	each disk. We found that the (dust) disks in our sample are much too small to have been significantly affected by tidal truncation. This could have a natural explanation, as the gas disk extent is expected to a $\sim$ a few times larger than the dust disk extent. On the other hand, if this is not the case, it could suggest 
	that the   
	smaller disk radii in the binary systems are primordial, rather than a product of binary interaction after disk formation. This may be counter-intuitive, because binary systems tend to have larger angular momenta than single systems. One possibility is that most of the angular
	momentum of a binary system is stored in the binary orbit, leaving less for the circumstellar disks. In any case, this is an intriguing result
	that disk and binary formation theories should seek to address.
	\item We detected several transition disks, two of which are the first ever millimeter observations (ROph3 = 2MASS J16230923-2417047; ROph 38 = WSB 82), 
	whereas one (ROph38 = WSB 82) is being classified as a transition disk for the first time based solely on the presence of a millimeter cavity unexpected from the available infrared data. In particular, WSB 82 is a transition disk with
	a noticeable gap in the low surface brightness outer disk that resembles the gaps seen in ALMA images of other Class II disks so far (e.g., HL Tau, 
	\citealt{al15}, HD 163296, \citealt{is16}, and TW Hya, \citealt{an16}). Interestingly, we find an intriguing trend that the transition disks are on average
	much brighter and larger than both Class I and Class II disks. Theoretical studies of disk evolution need to account for this trend.
	\item We have discovered an unexpected millimeter companion to the Class II source WLY 2-69 at 7.56$\farcs$ ($\sim 1000$ AU); 
	given the density of millimeter-wave background sources, it is most likely physically associated with the source. A search of the literature 
	on multiplicity in Oph yielded no reports of an optical or infrared companion. An examination of archival images from \textit{Spitzer}, however, shows this companion source in the infrared.
\end{itemize}

Due to the sheer number	of baselines available, as well as the very sensitive
receivers on the antennas, ALMA is producing exciting results almost daily,
particularly in the study of protoplanetary disks. Surveys such	as this one
of 49 targets in $\rho$ Ophiuchi as well as that of a set of 92 sources
in the $\sigma$ Ori cluster \citep{an17} demonstrate
conclusively that ALMA as a rapid survey instrument is coming	into its own.
\clearpage

\acknowledgments
{\centering ACKNOWLEDGMENTS \par}\
We thank the anonymous referee for their helpful comments.
EGC is supported by SOSPA3-017. ZYL is supported in part by NASA NNX 14AB38G and NSF AST-1313083 and 1716259.
This paper makes use of the following ALMA data:
ADS/JAO.ALMA\#
2013.1.00157.S. 
ALMA is a partnership of ESO (representing its member states), 
NSF (USA) and NINS (Japan), together with NRC (Canada), 
NSC and ASIAA (Taiwan), and KASI (Republic of Korea), in cooperation with the Republic of Chile. 
The Joint ALMA Observatory is operated by ESO, AUI/NRAO and NAOJ.
The National Radio Astronomy Observatory is a facility of the National Science Foundation operated under cooperative agreement by Associated Universities, Inc.


\end{document}